\documentclass[twocolumn]{aastex7}
\usepackage{natbib}
\usepackage{ulem}
\usepackage{xcolor}
\definecolor{BrickRed}{rgb}{0.8,0.1,0.1}

\begin{document}

\title{ALMA High-\emph{J} CO Spectroscopy of High-Redshift Galaxies. I. An Archive-based Catalog of CO Spectral Line Energy Distributions}

\author[orcid=0000-0001-9728-8909]{Ken-ichi Tadaki}
\affiliation{Faculty of Engineering, Hokkai-Gakuen University, Toyohira-ku, Sapporo 062-8605, Japan}
\email[show]{tadaki@hgu.jp}  


\begin{abstract}
High-$J$ CO emission in high-redshift galaxies has been studied primarily on an individual-source basis, limiting our ability to draw population-level conclusions about molecular-gas excitation.
To address this limitation, we present a catalog of CO spectroscopy based on archival data from the Atacama Large Millimeter/submillimeter Array (ALMA) for a sample of galaxies at $z > 3$, focusing on high-$J$ transitions ($J_{\rm up} = 9$--17).
Combining ALMA archival data with published measurements, we compile CO spectral line energy distributions (SLEDs) for 38 well-studied systems spanning $z \simeq 3.1$--6.9, including 5 hot dust-obscured galaxies (Hot DOGs), 17 submillimeter-bright galaxies (SMGs), and 16 optically selected quasars. 
The class-median SLEDs rise steeply to $J_{\rm up} = 9$ and remain approximately flat through $J_{\rm up} \sim 11$--12.
SMGs show relatively stronger low- to mid-$J$ emission relative to CO $J$=9--8, while Hot DOGs exhibit tentative evidence for higher excitation.
Comparison with simple excitation models suggests that X-ray dominated region (XDR) heating or dense, shock-heated gas can account for the extended high-$J$ CO SLEDs.
A tentative anti-correlation between the CO(9--8)-to-infrared luminosity ratio and excitation among the dusty galaxy populations suggests that the enhanced excitation in Hot DOGs may be driven by XDR heating from obscured AGN activity rather than by shocks.
Because the sample is restricted to sources with detected high-$J$ CO lines, it is biased toward highly excited systems. 
It nonetheless provides a reference for designing future systematic surveys of high-$J$ CO emission with a statistically complete or flux-limited sample.
\end{abstract}


\section{Introduction}\label{sec:intro}

Rotational transitions of CO are among the most widely used tracers of molecular gas in galaxies.
At high redshift, CO line observations directly constrain the gas reservoir that fuels intense star formation and black-hole growth, and they enable comparisons of interstellar medium (ISM) conditions across different galaxy populations.
While low-$J$ CO lines primarily trace the bulk of the cooler molecular gas reservoir, this work focuses on high-$J$ CO transitions ($J_{\rm up} \geq 9$), which have considerably higher excitation energies and critical densities.
For example, CO $J$=9--8 has an upper-level energy of $E_{\rm up}/k \simeq 249$~K and a critical density of order $10^{6}$~cm$^{-3}$, while CO(17--16) reaches $E_{\rm up}/k \simeq 846$~K (Table~\ref{tab:CO_transitions}).
Bright high-$J$ emission therefore requires ISM phases that are substantially warmer and/or denser than those responsible for the low-$J$ emission.

\begin{figure}[h]
\centering
\includegraphics[scale=1]{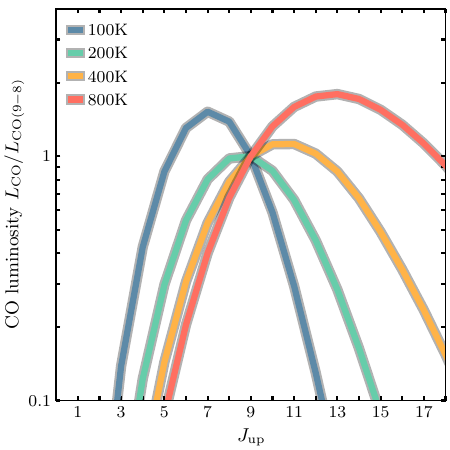}
\caption{
Normalized CO SLEDs predictedby non-LTE models. All curves are normalized to the CO(9--8) luminosity.
The calculations were performed with the \texttt{ndradex} package \citep{Taniguchi2023}, which implements the RADEX formalism \citep{2007AA...468..627V}, assuming an LVG geometry with $n_{\rm H_2}=10^{5}\,{\rm cm^{-3}}$, $N_{\rm CO}=10^{18}\,{\rm cm^{-2}}$, $\Delta v = 100~{\rm km~s^{-1}}$, and a background radiation temperature of $T_{\rm bg}=16.35$~K corresponding to the cosmic microwave background at $z=4$.
}
\label{fig:radex}
\end{figure}

Figure~\ref{fig:radex} shows normalized CO spectral line energy distributions (SLEDs) computed with non-LTE radiative transfer models.
As the kinetic temperature increases, the SLED peak shifts toward higher $J$ and the high-$J$ tail becomes more extended.
CO SLEDs therefore offer direct diagnostics of the excitation state of the molecular gas and of the relative importance of radiative and mechanical heating in the ISM.
Studies of nearby luminous galaxies have shown that SLED shapes vary widely and that high-$J$ emission often requires heating mechanisms beyond a single photodissociation region (PDR) component \citep[e.g.,][]{2015ApJ...801...72R,2015ApJ...802...81M}.
Strong far-ultraviolet radiation from compact starbursts heats gas in PDRs, while hard X-ray irradiation from an active galactic nucleus (AGN) can produce highly excited CO SLEDs in X-ray dominated regions \citep[XDRs; e.g.,][]{2010AA...518L..42V,2018MNRAS.474.3640M}.
Mechanical heating through turbulence, outflows, or merger-driven shocks offers an additional pathway to excite high-$J$ CO \citep[e.g.,][]{2013ApJ...762L..16M}.

\begin{table}[ht]
  \caption{CO Rotational Transitions from $J$=1--0 to $J$=17--16}
  \label{tab:CO_transitions}
  {\centering
  \begin{tabular}{lccc}
\hline
    Transition    & {Frequency} & {Excitation} & {Critical} \\
        & & {Potential} & {Density} \\
        & {(GHz)} & {(K)} & {(cm$^{-3}$)} \\
\hline
    CO $J$=1--0    & 115.27  & 5.5   & $2.2\times10^3$ \\
    CO $J$=2--1    & 230.54  & 16.6  & $1.0\times10^4$ \\
    CO $J$=3--2    & 345.80  & 33.2  & $3.1\times10^4$ \\
    CO $J$=4--3    & 461.04  & 55.3  & $7.6\times10^4$ \\
    CO $J$=5--4    & 576.27  & 83.0  & $1.9\times10^5$ \\
    CO $J$=6--5    & 691.47  & 116.2 & $3.7\times10^5$ \\
    CO $J$=7--6    & 806.65  & 154.9 & $5.9\times10^5$ \\
    CO $J$=8--7    & 921.80  & 199.1 & $8.1\times10^5$ \\
    CO $J$=9--8    & 1036.91 & 248.9 & $1.0\times10^6$ \\
    CO $J$=10--9   & 1151.98 & 304.2 & $1.3\times10^6$ \\
    CO $J$=11--10  & 1267.01 & 365.0 & $1.6\times10^6$ \\
    CO $J$=12--11  & 1381.99 & 431.3 & $1.9\times10^6$ \\
    CO $J$=13--12  & 1496.92 & 503.1 & $2.2\times10^6$ \\
    CO $J$=14--13  & 1611.79 & 580.5 & $2.6\times10^6$ \\
    CO $J$=15--14  & 1726.60 & 663.4 & $2.9\times10^6$ \\
    CO $J$=16--15  & 1841.35 & 751.7 & $3.5\times10^6$ \\
    CO $J$=17--16  & 1956.02 & 845.6 & $4.0\times10^6$ \\
\hline
  \end{tabular}\par}
  \vspace{1ex}
  \par\footnotesize
  \noindent\textbf{Notes:}
  Frequencies and upper-level energies are adopted from the CO entry in the Leiden Atomic and Molecular Database (LAMDA; \citealt{2005AA...432..369S}).
  Critical densities are computed at $T_{\rm kin}=10$~K using the two-level approximation with CO--p-H$_2$ collisional rates from LAMDA.
\end{table}

High-$J$ CO lines fall in the far-infrared regime in the rest frame and are therefore difficult to observe in the local Universe.
At high redshift, however, these transitions shift into the submillimeter atmospheric windows accessible to ALMA, enabling systematic studies of highly excited molecular gas in luminous galaxies.
To date, high-$J$ CO measurements at high redshift have largely been reported as targeted, source-by-source case studies \citep[e.g.,][]{2011ApJ...741L..37B,2012AA...545A..57S,2013Natur.496..329R,2014MNRAS.445.2848G,2015ApJ...808L...4A,2017AA...608A.144Y,2019ApJ...880....2W,2019AA...628A..23A,2019ApJ...880..153Y,2019MNRAS.489.3939C,2020ApJ...889..162L,2020MNRAS.494.5542R,2021ApJ...907...62R,2021ApJ...913..141R,2021ApJ...921...97J,2021ApJ...913...41L,2021AA...652A..66P,2022AA...662A..60D,2022PASJ...74.1429T,2023AA...674L...5B,2024MNRAS.535.1533B,2024AA...684A..56K,2024ApJ...977..190X,2024ApJ...962..119L,2025AA...703A.216H}.
Moving toward population-level statistical studies of molecular-gas excitation requires a larger number of objects with consistently measured CO SLEDs.
The ALMA Science Archive now contains a substantial and growing body of high-$J$ CO observations, but these data are spread across numerous projects and have not been compiled in a uniform format.

The primary goal of this paper is to provide a homogeneous catalog of high-$J$ CO measurements for galaxies at $z \gtrsim 3$.
This compilation is intended to serve as a reference for designing and interpreting future systematic surveys targeting statistically complete or flux-limited samples.
The paper is organized as follows.
Section~\ref{sec:sample} describes the ALMA archival search and sample definition.
Section~\ref{sec:reduction} describes the data reduction, imaging, spectral fitting, and flux measurements.
Section~\ref{sec:results} constructs CO SLEDs and characterizes the excitation properties of individual sources and of each galaxy class.
Section~\ref{sec:discussion} discusses the empirical excitation differences and places them in the context of simple PDR, XDR, and shock models.
Section~\ref{sec:summary} summarizes the main results.
In a companion paper (Paper~II; \citealt{2026arXiv260301352T}), we present 0\farcs03-resolution ALMA high-$J$ CO observations of one of the sources in this catalog, revealing spatially resolved XDR-like excitation gradients and providing a dynamical constraint on the central black hole mass.

\section{ALMA Archival Search}\label{sec:sample}

We do not attempt to construct a statistically complete or flux-limited sample, but rather assemble well-characterized sources for which the ALMA archive contains suitable CO spectroscopy.
We began by performing a project-level query of the ALMA Science Archive\footnote{\url{https://almascience.org/aq/}}.
To identify programs targeting highly excited molecular gas, we selected projects whose abstracts contain keywords related to high-$J$ CO observations (e.g., ``high-$J$ CO'' and ``CO(11--10)'').
In addition to explicit CO keywords, we searched for related emission lines, such as [N\,{\sc ii}]~205~$\mu$m and H$_2$O, to capture programs in which high-$J$ CO transitions may fall within the observed bandwidth.
From the projects returned by this query, we identified candidate targets with spectroscopic redshifts $z>3$.
This redshift cut ensures that even very highly excited transitions (e.g., CO(12--11)) are redshifted into ALMA Bands~7 and below, avoiding the highest-frequency bands where sensitivity and atmospheric transmission are less favorable.
For each candidate identified in the project-level search, we then queried the ALMA archive by sky position to retrieve all public datasets available as of February~1, 2026 that could cover CO rotational transitions, regardless of the original science goals.
We retained sources for which at least one high-$J$ CO line was clearly detected upon visual inspection of the archival data products.
While our primary focus is on high-$J$ CO ($J_{\rm up}\ge 9$), we also include mid-$J$ CO observations when available for the same targets.

The resulting target compilation is summarized in Table~\ref{tab:alma_summary}.
For the purpose of comparing CO excitation across different populations, we assign each target to one of three broad classes based on the selection criteria in the original discovery literature.
Hot Dust-Obscured Galaxies (Hot DOGs) are extreme, heavily obscured systems with unusually hot dust and very high infrared luminosities, often interpreted as hosting powerful dust-enshrouded AGN \citep{2012ApJ...755..173E,2015ApJ...804...27A,2015ApJ...805...90T}.
Submillimeter-bright galaxies (SMGs) refer to galaxies whose dust continuum emission is sufficiently bright to be detected with single-dish submillimeter telescopes or the \textit{Herschel} Space Observatory \citep[e.g.,][]{2008MNRAS.385.2225S,2013ApJ...767...88W,2016ApJ...832...78I}, and are therefore distinct from the much fainter ($S_{1{\rm mm}}< 1$~mJy) dusty star-forming galaxies that became accessible primarily through ALMA observations \citep[e.g.,][]{2013ApJ...769L..27H,2016ApJ...833...68A,2017MNRAS.466..861D}.
The quasar class comprises optically and/or near-infrared-selected quasars drawn primarily from wide-area surveys such as the Sloan Digital Sky Survey and the Canada--France--Hawaii Quasar Survey \citep[e.g.,][]{2013ApJ...779...24V,2015Natur.518..512W,2015ApJ...798...28K,2015ApJ...805L...8B,2016ApJ...833..222J}.
As a result, this subset is biased toward the most luminous, relatively unobscured quasars at high redshift and can be regarded as sampling the bright end of the quasar luminosity function \citep{2018ApJ...869..150M}.
This classification scheme is adopted to organize the catalog and to facilitate class-based comparisons of high-$J$ CO excitation within a consistent framework.


\section{Data Analysis}\label{sec:reduction}

\subsection{Continuum subtraction}\label{sec:reduction_contsub}

For all targets and projects listed in Table~\ref{tab:alma_summary}, we retrieved the calibrated visibility data primarily from the East Asian ALMA Regional Center (EA-ARC) data delivery service, supplemented where necessary with data from the ALMA Science Archive. 
Because the archived data were calibrated with the ALMA pipeline version current at the time of each delivery, the underlying pipeline versions can differ from project to project, as is standard for archival datasets. 
No additional manual calibration was performed.

Continuum subtraction was performed in the $uv$-plane using \texttt{CASA/uvcontsub\_old}.
For each dataset, the continuum was modeled as a zeroth-order polynomial (\texttt{fitorder=0}) over one or two contiguous sidebands.
Channels within the expected line window were excluded by masking a broad velocity range around the line center predicted from the literature redshift, with the remaining line-free channels used to determine the continuum level.
The continuum-subtracted visibilities were then used for imaging and subsequent line measurements.

\begin{deluxetable*}{lllccrccc}
\tablewidth{0pt}
\tablecaption{ALMA Observation Summary and Imaging Parameters \label{tab:alma_summary}}
\tablehead{
\colhead{Target} &
\colhead{$z$} &
\colhead{Transition} &
\colhead{Project ID} &
\colhead{Band} &
\colhead{$t_{\rm int}$} &
\colhead{MRS$^{\dagger}$} &
\colhead{$uv$ taper} &
\colhead{Beam Size}
}
\startdata
\multicolumn{9}{c}{Hot DOG sample} \\
\hline
W2305$-$0039 & 3.111 & CO(7--6)   & 2022.1.00353.S & 5 & 13 min & 13$''$   & --        & 2\farcs1$\times$1\farcs4 \\
            &  & CO(9--8)   & 2021.1.00168.S & 6 & 5 min  & 3\farcs4 & 1\farcs0  & 1\farcs2$\times$1\farcs1 \\
            &  & CO(11--10) & 2021.1.00168.S & 7 & 8 min  & 4\farcs5 & 1\farcs0  & 1\farcs3$\times$1\farcs2 \\
W0116$-$0505 & 3.190 & CO(7--6)   & 2022.1.00353.S & 5 & 17 min & 11$''$   & --        & 1\farcs6$\times$0\farcs9 \\
            &  & CO(9--8)   & 2021.1.00168.S & 6 & 5 min  & 3\farcs7 & 1\farcs0  & 1\farcs2$\times$1\farcs1 \\
            &  & CO(11--10) & 2021.1.00168.S & 7 & 8 min  & 3\farcs0 & 1\farcs0  & 1\farcs3$\times$1\farcs2 \\
W2246$-$7143 & 3.464 & CO(6--5)   & 2021.1.00168.S & 4 & 5 min  & 3\farcs5 & 1\farcs0  & 1\farcs3$\times$1\farcs1 \\
            &  & CO(9--8)   & 2021.1.00168.S & 6 & 5 min  & 4\farcs7 & 1\farcs0  & 1\farcs3$\times$1\farcs2 \\
            &  & CO(11--10) & 2021.1.00168.S & 7 & 8 min  & 4\farcs1 & 1\farcs0  & 1\farcs3$\times$1\farcs2 \\
W0831+0140 & 3.914 & CO(7--6)   & 2022.1.00353.S & 5 & 11 min & 11$''$   & --        & 1\farcs2$\times$1\farcs1 \\
           &  & CO(10--9)  & 2021.1.00168.S & 6 & 5 min  & 5\farcs1 & 1\farcs0  & 1\farcs3$\times$1\farcs2 \\
W2246$-$0526 & 4.601 & CO(9--8)   & 2021.1.00168.S & 5 & 12 min & 5\farcs4 & 0\farcs8  & 1\farcs2$\times$1\farcs1 \\
            &  & CO(11--10) & 2021.1.00168.S & 6 & 5 min  & 3\farcs8 & 0\farcs8  & 1\farcs2$\times$1\farcs1 \\
\hline
\multicolumn{9}{c}{SMG sample} \\
\hline
G12v2.30 & 3.260 & CO(9--8)   & 2015.1.01042.S & 6 & 26 min  & 5\farcs5    & 4\farcs0  & 3\farcs9$\times$3\farcs6 \\
         &       & CO(10--9)  & 2018.1.00966.S & 6 & 12 min & 4\farcs1    & 4\farcs0  & 4\farcs3$\times$3\farcs8 \\
NCv1.143 & 3.565 & CO(9--8)   & 2018.1.00966.S & 6 & 22 min  & 5\farcs4    & 4\farcs0  & 4\farcs0$\times$3\farcs8 \\
         &       & CO(10--9)  & 2018.1.00966.S & 6 & 34 min & 4\farcs6    & 4\farcs0  & 3\farcs8$\times$3\farcs7 \\
G09v1.97 & 3.634 & CO(9--8)   & 2019.1.00533.S & 6 & 32 min  & 6\farcs1    & 4\farcs0  & 4\farcs0$\times$3\farcs7 \\
         &       & CO(10--9)  & 2022.1.00346.S & 6 & 48 min & 7\farcs9    & 4\farcs0  & 3\farcs9$\times$3\farcs9 \\
G15v2.779 & 4.243 & CO(7--6)   & 2018.1.00861.S & 4 & 44 min  & 2\farcs9    & 3\farcs0  & 4\farcs2$\times$3\farcs8 \\
          &       & CO(9--8)   & 2018.1.00966.S & 6 & 24 min & 6\farcs4    & 4\farcs0  & 4\farcs2$\times$3\farcs8 \\
          &       & CO(10--9)  & 2018.1.00966.S & 5 & 62 min & 7\farcs1    & 4\farcs0  & 4\farcs0$\times$3\farcs8 \\
SPT0544$-$40 & 4.269 & CO(4--3)   & 2015.1.00504.S & 3 & 1.2 min & 29$''$    & --  & 4\farcs8$\times$2\farcs8 \\
            &  & CO(5--4)    & 2015.1.00504.S & 3 & 1.2 min & 23$''$    & --  & 3\farcs2$\times$2\farcs6 \\
            &  & CO(10--9)   & 2023.1.01532.S & 6 & 5 min & 12$''$    & 3\farcs0 & 3\farcs3$\times$3\farcs2 \\
SPT2311$-$54 & 4.278 & CO(4--3)   & 2012.1.00844.S & 3 & 3 min & 26$''$    & --  & 4\farcs2$\times$2\farcs9 \\  
            &  & CO(5--4)    & 2012.1.00844.S & 3 & 1.5 min & 18$''$    & --  & 3\farcs4$\times$2\farcs0 \\
            &  & CO(10--9)   & 2023.1.01532.S & 6 & 24 min & 12$''$    & 3\farcs0 & 3\farcs4$\times$3\farcs2 \\
SPT2349$-$56N1 & 4.303 & CO(7--6)   & 2021.1.01313.S & 4 & 44 min & 7$''$    & 1\farcs0  & 1\farcs5$\times$1\farcs2 \\
            &  & CO(11--10) & 2021.1.01313.S & 6 & 39 min & 8$''$    & 1\farcs0  & 1\farcs5$\times$1\farcs4 \\
COS-AzTEC-1 & 4.342 & CO(7--6)   & 2017.A.00032.S & 4 & 27 min & 9\farcs6 & 0\farcs5  & 1\farcs2$\times$1\farcs1 \\
            &  & CO(12--11) & 2018.1.00081.S & 6 & 296 min& 4\farcs7 & 1\farcs0  & 1\farcs3$\times$1\farcs2 \\
BR1202$-$0725 & 4.693 & CO(10--9) & 2017.1.01516.S & 5 & 70 min & 7\farcs5 & 0\farcs5  & 1\farcs1$\times$1\farcs1 \\
             &  & CO(13--12)& 2017.1.00963.S & 4 & 25 min & 10$''$   & --        & 1\farcs3$\times$1\farcs1 \\
             &  & CO(14--13)& 2017.1.00963.S & 4 & 22 min & 9\farcs3 & --        & 1\farcs6$\times$1\farcs0 \\
COS-AzTEC-3 & 5.298 & CO(12--11) & 2015.1.00928.S & 6 & 25 min & 11$''$   & --        & 1\farcs9$\times$1\farcs4 \\
            &  & CO(13--12) & 2016.1.00330.S & 6 & 30 min & 12$''$   & --        & 1\farcs9$\times$1\farcs7 \\
            &  & CO(14--13) & 2016.1.00330.S & 6 & 30 min & 12$''$   & --        & 1\farcs7$\times$1\farcs5 \\
G09-83808 & 6.027 & CO(10--9) & 2023.1.01281.S & 5 & 290 min & 12$''$ & 1\farcs5 & 2\farcs3$\times$2\farcs0 \\
SPT0311$-$58 & 6.903 & CO(11--10)& 2021.1.01015.S & 5 & 93 min & 5\farcs6 & 0\farcs8  & 1\farcs2$\times$1\farcs1 \\
             &  & CO(12--11)& 2021.1.01015.S & 5 & 93 min & 5\farcs6 & 0\farcs8  & 1\farcs2$\times$1\farcs1 \\
\enddata
\end{deluxetable*}

\setcounter{table}{1}

\begin{deluxetable*}{lllccrccc}
\tablewidth{0pt}
\tablecaption{ALMA Observation Summary and Imaging Parameters (continued)}
\tablehead{
\colhead{Target} &
\colhead{$z$} &
\colhead{Transition} &
\colhead{Project ID} &
\colhead{Band} &
\colhead{$t_{\rm int}$} &
\colhead{MRS$^{\dagger}$} &
\colhead{$uv$ taper} &
\colhead{Beam Size}
}
\startdata
\multicolumn{9}{c}{Quasar sample} \\
\hline
J2310+1855 & 6.003 & CO(13--12) & 2019.1.00080.S & 6 & 222 min& 0\farcs9 &  1\farcs0  & 1\farcs1$\times$0\farcs8 \\
           &  & CO(14--13) & 2019.1.00080.S & 6 & 222 min& 0\farcs9 &  1\farcs0 & 1\farcs1$\times$0\farcs8 \\
J2219+0102   & 6.150 & CO(7--6)  & 2019.1.00147.S & 3 & 52 min & 13 & --        & 1\farcs3$\times$1\farcs1 \\
&  & CO(9--8)  & 2023.1.00852.S & 4 & 55 min & 7\farcs3 & 0\farcs6  & 1\farcs2$\times$1\farcs2 \\
J0100+2802 & 6.327 & CO(12--11) & 2023.1.00653.S & 5 & 61 min & 8\farcs3 & 0\farcs8   & 1\farcs3$\times$1\farcs1 \\
           &  & CO(16--15) & 2019.1.00466.S & 6 & 150 min& 12$''$        & --        & 1\farcs6$\times$1\farcs3 \\
           &  & CO(17--16) & 2019.1.00466.S & 6 & 150 min& 12$''$        & --        & 1\farcs5$\times$1\farcs2 \\
PJ036+03     & 6.541 & CO(9--8)  & 2023.1.00852.S & 4 & 44 min & 5\farcs7 & 0\farcs8  & 1\farcs2$\times$1\farcs2 \\
             &.      & CO(10--9) & 2023.1.00852.S & 4 & 44 min & 5\farcs7 & 0\farcs8  & 1\farcs2$\times$1\farcs1 \\
J2348$-$3054 & 6.902 & CO(9--8)  & 2023.1.00852.S & 4 & 39 min & 4\farcs3 & 0\farcs8  & 1\farcs1$\times$1\farcs0 \\
             &.      & CO(10--9) & 2023.1.00852.S & 4 & 39 min & 4\farcs3 & 0\farcs8  & 1\farcs1$\times$1\farcs0 \\
\enddata
\tablecomments{$^{\dagger}$Maximum recoverable scale.}
\end{deluxetable*}


\subsection{Imaging strategy}\label{sec:reduction_imaging}

Imaging was performed with \texttt{CASA/tclean}.
We used Briggs weighting with a common \texttt{robust} parameter of $+2.0$ for all datasets.
The $uv$ taper was adjusted on a dataset-by-dataset basis.
The spectral cubes were constructed with a channel width of 100~km~s$^{-1}$.
We set \texttt{niter=0} and produced dirty cubes for the subsequent flux measurements.

\begin{figure}
  \centering
  \includegraphics[scale=1.0]{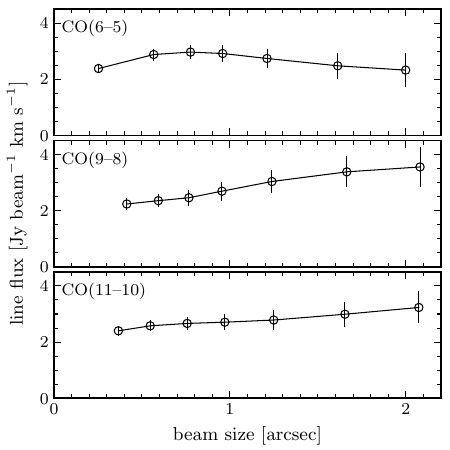}
  \caption{
Growth-curve test for W2246$-$7143. 
Integrated CO line flux as a function of synthesized beam size, measured by re-imaging the calibrated visibility data with a range of $uv$ tapers.
  }
  \label{fig:growthcurve_W2246-7143}
\end{figure}

The archival datasets assembled for this work span a wide range of angular resolutions ($\sim0\farcs2$--$3\farcs0$ prior to tapering), reflecting the heterogeneous array configurations across projects.
Since our primary goal is to measure spatially integrated CO line fluxes, we adopted an imaging strategy designed to recover the total emission without incurring unnecessary sensitivity losses.
To empirically determine the beam size required to encompass the bulk of the emission, we performed a growth-curve test using W2246$-$7143, for which multiple CO transitions are available (CO(6--5), CO(9--8), and CO(11--10)).
We re-imaged the same calibrated visibility data with a sequence of $uv$ tapers to produce a range of synthesized beam sizes and measured the line flux as a function of beam size (Figure~\ref{fig:growthcurve_W2246-7143}).
The measured line flux increases with beam size and converges to an approximately constant level by $\sim1''$.
Although a mild further increase is apparent toward $\sim2''$ for some transitions, the change beyond $\sim1''$ is within the measurement uncertainties and is therefore not statistically significant.
We thus adopt a baseline imaging resolution of $\sim1''$, which captures the bulk of the CO emission without the sensitivity penalty of further tapering.

For the majority of the datasets used in this work, the maximum recoverable scale (MRS) of the CO observations is typically $\gtrsim 3''$.
This comfortably exceeds the angular scales relevant to our flux measurements, so significant spatial filtering of the total CO emission is not expected for most targets.
One notable exception is J2310+1855, for which the MRS is only $\simeq 0\farcs9$.
We therefore caution that its integrated CO flux may be underestimated if a non-negligible fraction of the emission extends beyond the MRS.

In addition, several gravitationally lensed systems exhibit CO emission extending beyond $\sim 1''$, in some cases resolved into multiple components.
These cases require individual treatment as follows.
For SPT0311$-$58W, we measure the CO line flux from an image smoothed with \texttt{CASA/imsmooth} to a $1\farcs5 \times 1\farcs0$ beam to encompass the extended structure.
For G12v2.30, NCv1.143, G09v1.97, G15v2.779, SPT0544$-$40, and SPT2311$-$54, we adopt images produced with a $uv$ taper to achieve synthesized beams larger than $\sim3$--$4''$.
For G09--83808, we measure the CO line flux as the sum of the two lensed components that are spatially separated in the ALMA maps.
We note that CO line detections in higher-resolution images typically achieve higher per-beam signal-to-noise ratios (S/N) than the S/N of the total integrated fluxes reported in this work.
The adopted tapers and resulting synthesized beams are summarized in Table~\ref{tab:alma_summary}.

\subsection{Line-profile fitting}\label{sec:reduction_spec}

The line-profile fitting procedure serves three purposes: (i) to define an appropriate velocity integration range for measuring galaxy-integrated CO line fluxes, (ii) to correct the measured CO fluxes for blending by nearby transitions, and (iii) to apply a model-based correction when the spectral coverage does not fully span the required velocity interval.

To carry out this fitting, we extract the source spectrum at the peak position of the moment-0 map (Section \ref{sec:reduction_flux}). 
The peak position and the velocity integration range are found iteratively, by remaking the moment-0 map over the velocity range from the line fit and re-locating the peak until both converge.
For each source, we model the extracted spectra as Gaussian line profiles, fitting all available transitions simultaneously.
We assume a common centroid shift relative to the systemic redshift and a common line width for all fitted lines of a given source, while allowing the peak amplitude of each transition to vary independently.
When nearby transitions may contaminate the target CO line, we include additional Gaussian components in the same fit.
The most relevant cases are [C\,{\sc i}]($^3P_2\!\rightarrow{}^3P_1$) near CO(7--6) and H$_2$O $3_{1,2}$--$2_{2,1}$ near CO(10--9), the latter being separated from CO(10--9) by only $\sim300~\mathrm{km~s^{-1}}$.
The fitting is performed with a non-linear least-squares optimizer (SciPy \texttt{least\_squares}) using a weighted residual formulation.
For each spectrum, the rms noise is estimated from line-free channels by excluding a broad velocity range around the expected centers of the CO line and any relevant contaminant lines.
Residuals are then normalized by this per-spectrum rms, so that noisier spectra receive less weight in the joint solution.

\begin{figure}
  \centering
  \includegraphics[scale=1.0]{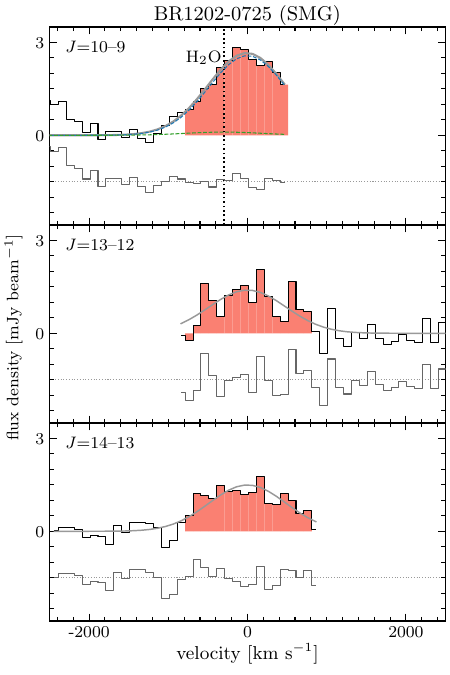}
\caption{
CO(10--9), CO(13--12), and CO(14--13) spectra of BR1202$-$0725 (SMG component).
Velocities are referenced to the best-fit CO line center.
In each panel the black histogram is the observed spectrum and the gray solid curve is the total composite model. 
In the top panel the blue and green dashed curves show the CO and H$_2$O components of the fit, and the vertical dotted line marks the expected position of H$_2$O.
Shaded channels indicate the velocity range adopted for the integrated-flux measurements.
Below each spectrum we show the residual after subtracting the total model, in the same flux units and offset so that a zero residual lies on the horizontal dotted line.
}
  \label{fig:spectra_BR1202-SMG}
\end{figure}

We define the velocity integration range for each CO line based on the best-fit line profile.
Specifically, we select all spectral channels in which the best-fit  Gaussian profile exceeds 30\% of its peak value.
This threshold balances total flux recovery against S/N: a narrower range risks missing part of the galaxy-integrated emission, while a wider range includes the low-level wings where the flux density is small, adding noise and degrading the S/N of the integrated flux.
Because the wings below this threshold are excluded, the measured flux can slightly underestimate the total line flux.
From the best-fit Gaussian model we find that the adopted velocity window contains on average a fraction of $0.93 \pm 0.02$ of the total Gaussian flux across all CO lines in our sample. 
The same velocity range is applied to every transition of a given source, and the CO SLED is normalized to the CO(9-8) luminosity (Section \ref{sec:results}), so this common factor cancels in the line ratios and the SLED is essentially unaffected.

Within the CO integration range $(v_{\rm min},v_{\rm max})$ defined above, we estimate the fractional contribution of each contaminant line from the best-fit models.
For example, for H$_2$O contaminating CO(10--9) we define
\begin{equation}
f_{\rm H_2O} =
\frac{\int_{v_{\rm min}}^{v_{\rm max}} S_{\rm H_2O}(v)\,dv}
     {\int_{v_{\rm min}}^{v_{\rm max}}
     \left[S_{\rm CO}(v)+S_{\rm H_2O}(v)\right]\,dv},
\end{equation}
and compute an analogous fraction for [C\,{\sc i}] in the CO(7--6) spectral window.

When the observed bandwidth does not fully cover the intended integration interval, we estimate a correction factor from the best-fit Gaussian profile.
We define the fraction of the model CO flux captured within the actually observed velocity range, $f_{\rm cov}$, as
\begin{equation}
f_{\rm cov} =
\frac{\int_{v_{\rm cov}} S_{\rm CO}(v)\,dv}
     {\int_{v_{\rm min}}^{v_{\rm max}} S_{\rm CO}(v)\,dv},
\end{equation}
where $v_{\rm cov}$ is the velocity interval actually observed within $(v_{\rm min},v_{\rm max})$.

As an example, Figure~\ref{fig:spectra_BR1202-SMG} shows the CO(10--9), CO(13--12), and CO(14--13) spectra of BR1202$-$0725 (SMG) together with the best-fit models.
In the CO(10--9) panel, the nearby H$_2$O $3_{1,2}$--$2_{2,1}$ transition is explicitly included in the fit, enabling a deblended CO(10--9) flux measurement.
Within the CO(10--9) integration window, the modeled H$_2$O fraction is $f_{\rm H_2O}=0.038$, indicating that blending is a small but non-negligible effect.
We remove this contamination by subtracting the best-fit H$_2$O component.
In addition, the CO(10--9) spectrum does not fully cover the adopted integration range.
The fraction of the best-fit CO(10--9) profile covered by the observed channels is $f_{\rm cov}=0.891$, so we apply a correction factor of $1/f_{\rm cov}=1.122$ to recover the total CO(10--9) flux implied by the fitted line profile.
The corresponding spectra and best-fit models for all other sources analyzed in this work are presented in Appendix~\ref{app:spectra}.

For the gravitationally lensed sources G12v2.30, G09v1.97, G15v2.779, and G09$-$83808, a single-Gaussian profile does not adequately separate the H$_2$O contribution in the blended CO(10--9)+H$_2$O region, because differential magnification of spatially distinct emission components produces an asymmetric line shape.
We therefore model each line as a double-Gaussian profile representing the blue- and red-shifted lensed components, in which the two components share a common line width but are allowed to differ in amplitude.
The same double-Gaussian shape is imposed for both CO(10--9) and H$_2$O, so that the full model in the blended spectral region consists of four Gaussian components.
Beyond this line-profile effect, differential lensing can also bias the excitation we infer, because it preferentially magnifies the compact, highly excited gas relative to the more extended low-excitation gas \citep{2012MNRAS.424.2429S}.

\subsection{CO line flux measurements}\label{sec:reduction_flux}

For each CO transition, we constructed a moment-0 (velocity-integrated) map by integrating the spectral cube over the velocity interval determined from the line-profile fitting described in Section~\ref{sec:reduction_spec}.
We then adopted the peak value in the moment-0 map as the total CO line flux of each galaxy, $S_{\rm CO}\Delta v$.
Because the synthesized beam is sufficiently large to encompass the entire source in all cases, the CO emission is effectively unresolved, and the peak value in the moment-0 map (Jy\,beam$^{-1}$\,km\,s$^{-1}$) is equivalent to the spatially integrated line flux (Jy\,km\,s$^{-1}$).

The statistical uncertainty on $S_{\rm CO}\Delta v$ was estimated from the rms measured in source-free regions of the moment-0 map.
Because all targets are located at the phase center, no primary-beam correction was applied.
The uncertainties quoted in this paper are statistical only.
We note that the absolute flux calibration uncertainty of ALMA is typically $\lesssim$5\% in Bands~3--5 and $\lesssim$10\% in Bands~6--7. These systematic uncertainties are not included in the error budget.

The measured CO line fluxes are summarized separately for each source class.
Measurements for the Hot DOG sample are presented in Table~\ref{tab:co_flux_lum_hotdogs_thiswork}, those for the SMG sample in Table~\ref{tab:co_flux_lum_SMG_thiswork}, and those for the quasar sample in Table~\ref{tab:co_flux_lum_quasar_thiswork}.
For all sources listed in these tables, the redshifts are determined directly from our CO line fitting rather than adopted from the literature, to ensure internal consistency between the line measurements and the derived luminosities.

\begin{deluxetable*}{lclccc}
\tablewidth{0pt}
\tablecaption{CO line fluxes and luminosities for the Hot DOG sample analyzed in this work. \label{tab:co_flux_lum_hotdogs_thiswork}}
\tablehead{
\colhead{Target} & \colhead{$z$} & \colhead{CO transition} & \colhead{Flux} & \colhead{Luminosity} & \colhead{Reference}\\
 &  &  & \colhead{[Jy km s$^{-1}$]} & \colhead{[$L_\odot$]} & }
\startdata
W2305--0039 & 3.111 & CO(1--0)  & 0.025$\pm$0.003 & $(5.1\pm0.6)\times10^{5}$ & \citet{2020MNRAS.496.1565P}\\
            &       & CO(4--3)  & 2.70$\pm$0.19  & $(2.2\pm0.2)\times10^{8}$ & \citet{2024MNRAS.534..978M}\\
            &       & CO(7--6)  & 4.11$\pm$0.23  & $(5.9\pm0.3)\times10^{8}$ & this work\\
            &       & CO(9--8)  & 2.48$\pm$0.49  & $(4.6\pm0.9)\times10^{8}$ & this work\\
            &       & CO(11--10)& 3.44$\pm$0.32  & $(7.8\pm0.7)\times10^{8}$ & this work\\
W0116--0505 & 3.189 & CO(4--3)  & 0.76$\pm$0.07  & $(6.5\pm0.6)\times10^{7}$ & \citet{2024MNRAS.534..978M}\\
            &       & CO(7--6)  & 1.37$\pm$0.10  & $(2.1\pm0.2)\times10^{8}$ & this work\\
            &       & CO(9--8)  & 1.47$\pm$0.30  & $(2.8\pm0.6)\times10^{8}$ & this work\\
            &       & CO(11--10)& 0.99$\pm$0.31  & $(2.3\pm0.7)\times10^{8}$ & this work\\
W2246--7143 & 3.465 & CO(4--3)  & 1.57$\pm$0.12  & $(1.5\pm0.1)\times10^{8}$ & \citet{2024MNRAS.534..978M}\\
            &       & CO(6--5)  & 2.73$\pm$0.35  & $(4.0\pm0.5)\times10^{8}$ & this work\\
            &       & CO(9--8)  & 3.03$\pm$0.41  & $(6.7\pm0.9)\times10^{8}$ & this work\\
            &       & CO(11--10)& 2.79$\pm$0.35  & $(7.5\pm0.9)\times10^{8}$ & this work\\
W0831+0140  & 3.913 & CO(1--0)  & 0.036$\pm$0.003 & $(1.1\pm0.1)\times10^{6}$ & \citet{2020MNRAS.496.1565P}\\
            &       & CO(4--3)  & 2.85$\pm$0.17  & $(3.4\pm0.2)\times10^{8}$ & \citet{2024MNRAS.534..978M}\\
            &       & CO(7--6)  & 4.04$\pm$0.18  & $(8.4\pm0.4)\times10^{8}$ & this work\\
            &       & CO(10--9) & 4.63$\pm$0.37  & $(1.4\pm0.1)\times10^{9}$ & this work\\
W2246--0526 & 4.601 & CO(2--1)  & 0.35$\pm$0.03  & $(2.7\pm0.2)\times10^{7}$ & \citet{2025AA...703A.216H}\\
            &       & CO(5--4)  & 1.21$\pm$0.09  & $(2.3\pm0.2)\times10^{8}$ & \citet{2024MNRAS.534..978M}\\
            &       & CO(7--6)  & 2.20$\pm$0.03  & $(5.9\pm0.1)\times10^{8}$ & \citet{2025AA...703A.216H}\\
            &       & CO(9--8)  & 2.10$\pm$0.45  & $(7.2\pm1.6)\times10^{8}$ & this work\\
            &       & CO(11--10)& 2.13$\pm$0.23  & $(9.0\pm1.0)\times10^{8}$ & this work\\
            &       & CO(12--11)& 2.45$\pm$0.06  & $(1.1\pm0.03)\times10^{9}$ & \citet{2025AA...703A.216H}\\
\enddata
\end{deluxetable*}

\begin{deluxetable*}{lclccc}
\tablewidth{0pt}
\tablecaption{CO line fluxes and luminosities for the SMG sample analyzed in this work.\label{tab:co_flux_lum_SMG_thiswork}}
\tablehead{
\colhead{Target} & \colhead{$z$} & \colhead{CO transition} & \colhead{Flux} & \colhead{Luminosity} & \colhead{Reference}\\
 &  &  & \colhead{[Jy km s$^{-1}$]} & \colhead{[$L_\odot$]} & }
\startdata
G12v2.30$^{\dagger}$   & 3.260 & CO(4--3)  & 16.4$\pm$2.7 & $(1.5\pm0.2)\times10^{9}$ & \citet{2017AA...608A.144Y} \\
                       &       & CO(5--4)  & 22.6$\pm$2.0 & $(2.5\pm0.2)\times10^{9}$ & \citet{2017AA...608A.144Y} \\
                       &       & CO(6--5)  & 19.8$\pm$1.1 & $(2.6\pm0.2)\times10^{9}$ & \citet{2017AA...608A.144Y} \\
                       &       & CO(8--7)  & 20.7$\pm$1.8 & $(3.7\pm0.3)\times10^{9}$ & \citet{2017AA...608A.144Y} \\
                       &       & CO(9--8)  & 10.91$\pm$1.70 & $(2.2\pm0.3)\times10^{9}$ & this work \\
                       &       & CO(10--9) & 7.84$\pm$1.65 & $(1.7\pm0.4)\times10^{9}$ & this work \\
                       &       & CO(11--10) & 12.1$\pm$2.5 & $(3.0\pm0.6)\times10^{9}$ & \citet{2017AA...608A.144Y} \\
NCv1.143$^{\dagger}$   & 3.565 & CO(3--2)  &  5.5$\pm$0.6 & $(4.2\pm0.5)\times10^{8}$ & \citet{2017AA...608A.144Y} \\
                       &       & CO(5--4)  & 10.7$\pm$0.9 & $(1.4\pm0.1)\times10^{9}$ & \citet{2017AA...608A.144Y} \\
                       &       & CO(6--5)  &  9.6$\pm$1.2 & $(1.5\pm0.2)\times10^{9}$ & \citet{2017AA...608A.144Y} \\
                       &       & CO(7--6)  & 11.5$\pm$0.9 & $(2.1\pm0.2)\times10^{9}$ & \citet{2017AA...608A.144Y} \\
                       &       & CO(9--8)  & 7.51$\pm$0.91 & $(1.7\pm0.2)\times10^{9}$ & this work \\
                       &       & CO(10--9) & 6.53$\pm$0.67 & $(1.7\pm0.2)\times10^{9}$ & this work \\
G09v1.97$^{\dagger}$   & 3.632 & CO(3--2)  &  5.7$\pm$2.2 & $(4.5\pm1.7)\times10^{8}$ & \citet{2017AA...608A.144Y} \\
                       &       & CO(5--4)  &  9.7$\pm$1.2 & $(1.3\pm0.2)\times10^{9}$ & \citet{2017AA...608A.144Y} \\
                       &       & CO(6--5)  & 10.0$\pm$1.7 & $(1.6\pm0.3)\times10^{9}$ & \citet{2017AA...608A.144Y} \\
                       &       & CO(7--6)  &  8.0$\pm$0.9 & $(1.5\pm0.2)\times10^{9}$ & \citet{2017AA...608A.144Y} \\
                       &       & CO(9--8)  & 9.82$\pm$0.89 & $(2.3\pm0.2)\times10^{9}$ & this work \\
                       &       & CO(10--9) & 6.55$\pm$0.62 & $(1.7\pm0.2)\times10^{9}$ & this work \\
G15v2.779$^{\dagger}$  & 4.244 & CO(4--3)  & 7.5$\pm$0.9 & $(1.0\pm0.1)\times10^{9}$ & \citet{2011ApJ...740...63C} \\
                       &       & CO(5--4)  & 13.0$\pm$1.6 & $(2.2\pm0.3)\times10^{9}$ & \citet{2011ApJ...740...63C} \\
                       &       & CO(7--6)  & 8.13$\pm$1.74 & $(1.9\pm0.4)\times10^{9}$ & this work \\
                       &       & CO(9--8)  & 7.17$\pm$0.52 & $(2.2\pm0.2)\times10^{9}$ & this work \\
                       &       & CO(10--9) & 6.44$\pm$0.72 & $(2.2\pm0.2)\times10^{9}$ & this work \\
SPT0544$-$40$^{\dagger}$ & 4.269 & CO(4--3)  & 4.95$\pm$0.42 & $(6.8\pm0.6)\times10^{8}$ & this work \\
                       &       & CO(5--4)  & 7.02$\pm$0.39 & $(1.2\pm0.1)\times10^{9}$ & this work\\
                       &       & CO(7--6)  & 5.16$\pm$0.93 & $(1.2\pm0.2)\times10^{9}$ & \citet{2023AA...676A..89G}\\
                       &       & CO(10--9) & 4.16$\pm$0.31 & $(1.4\pm0.1)\times10^{9}$ & this work\\
SPT2311$-$54$^{\dagger}$ & 4.280 & CO(4--3)  & 3.21$\pm$0.24 & $(4.4\pm0.3)\times10^{8}$ & this work \\
                       &       & CO(5--4)  & 4.79$\pm$0.39 & $(8.2\pm0.7)\times10^{8}$ & this work\\
                       &       & CO(7--6)  & 4.72$\pm$0.61 & $(1.1\pm0.2)\times10^{9}$ & \citet{2023AA...676A..89G}\\
                       &       & CO(10--9) & 2.69$\pm$0.15 & $(9.2\pm0.5)\times10^{8}$ & this work\\
SPT2349$-$56N1   & 4.323 & CO(4--3)  & 1.55$\pm$0.03 & $(2.2\pm0.04)\times10^{8}$ & \citet{2020MNRAS.495.3124H} \\
                       &       & CO(7--6)  & 2.04$\pm$0.09 & $(5.0\pm0.2)\times10^{8}$ & this work\\
                       &       & CO(11--10)& 1.34$\pm$0.10 & $(5.1\pm0.4)\times10^{8}$ & this work\\
COS-AzTEC-1     & 4.342 & CO(4--3)  & 1.75$\pm$0.24 & $(2.5\pm0.3)\times10^{8}$ & \citet{2015MNRAS.454.3485Y} \\
                       &       & CO(5--4)  & 1.55$\pm$0.22 & $(2.7\pm0.4)\times10^{8}$ & \citet{2015MNRAS.454.3485Y} \\
                       &       & CO(7--6)  & 1.81$\pm$0.09 & $(4.4\pm0.2)\times10^{8}$ & this work\\
                       &       & CO(12--11)& 0.89$\pm$0.04 & $(3.7\pm0.2)\times10^{8}$ & this work\\
\enddata
\end{deluxetable*}

\setcounter{table}{3}

\begin{deluxetable*}{lclccc}
\tablewidth{0pt}
\tablecaption{CO line fluxes and luminosities for the SMG sample analyzed in this work. (continued)}
\tablehead{
\colhead{Target} & \colhead{$z$} & \colhead{CO transition} & \colhead{Flux} & \colhead{Luminosity} & \colhead{Reference}\\
 &  &  & \colhead{[Jy km s$^{-1}$]} & \colhead{[$L_\odot$]} & }
\startdata
BR1202$-$0725   & 4.686 & CO(2--1)  & 0.39$\pm$0.08 & $(3.1\pm0.6)\times10^{7}$ & \citet{2002AJ....123.1838C} \\
                &       & CO(5--4)  & 2.6$\pm$0.4   & $(5.1\pm0.8)\times10^{8}$ & \citet{2012AA...545A..57S} \\
                &       & CO(7--6)  & 3.1$\pm$0.4   & $(8.6\pm1.1)\times10^{8}$ & \citet{2012AA...545A..57S}\\
                &       & CO(10--9) & 2.81$\pm$0.13 & $(1.1\pm0.05)\times10^{9}$ & this work\\
                &       & CO(11--10)& 2.9$\pm$1.8   & $(1.3\pm0.8)\times10^{9}$ & \citet{2012AA...545A..57S}\\
                &       & CO(12--11)& 2.03$\pm$0.09 & $(9.6\pm0.4)\times10^{8}$ & \citet{2021ApJ...913...41L}\\
                &       & CO(13--12)& 1.58$\pm$0.30 & $(8.1\pm1.5)\times10^{8}$ & this work\\
                &       & CO(14--13)& 1.75$\pm$0.20 & $(9.7\pm1.1)\times10^{8}$ & this work\\
COS-AzTEC-3     & 5.297 & CO(2--1)  & 0.20$\pm$0.02 & $(1.9\pm0.2)\times10^{7}$ & \citet{2020ApJ...895...81R}\\
                &       & CO(5--4)  & 0.97$\pm$0.09 & $(2.3\pm0.2)\times10^{8}$ & \citet{2020ApJ...895...81R}\\
                &       & CO(6--5)  & 1.36$\pm$0.19 & $(3.9\pm0.5)\times10^{8}$ & \citet{2020ApJ...895...81R}\\
                &       & CO(12--11)& 0.77$\pm$0.09 & $(4.4\pm0.5)\times10^{8}$ & this work\\
                &       & CO(13--12)& 0.58$\pm$0.08 & $(3.6\pm0.5)\times10^{8}$ & this work\\
                &       & CO(14--13)& 0.49$\pm$0.09 & $(3.3\pm0.6)\times10^{8}$ & this work\\
G09-83808$^{\dagger}$ & 6.024 & CO(2--1) & 0.6$\pm$0.1 & $(6.9\pm1.1)\times10^{7}$ & \citet{2022ApJ...933..242Z}\\
                      &       & CO(5--4) & 0.92$\pm$0.30 & $(2.6\pm0.9)\times10^{8}$ & \citet{2017MNRAS.472.2028F}\\
                      &       & CO(6--5) & 0.87$\pm$0.24 & $(3.0\pm0.8)\times10^{8}$ & \citet{2017MNRAS.472.2028F}\\
                      &       & CO(10--9) & 0.61$\pm$0.02 & $(3.5\pm0.1)\times10^{8}$ & this work\\
                      &       & CO(12--11)& 0.49$\pm$0.09 & $(3.4\pm0.6)\times10^{8}$ & \citet{2022PASJ...74.1429T}\\
SPT0311$-$58W$^{\dagger}$   & 6.906 & CO(6--5)  & 2.33$\pm$0.09 & $(9.8\pm0.4)\times10^{8}$ & \citet{2021ApJ...921...97J}\\
                &       & CO(7--6)  & 2.15$\pm$0.12 & $(1.1\pm0.06)\times10^{9}$ & \citet{2021ApJ...921...97J}\\
                &       & CO(10--9) & 1.41$\pm$0.10 & $(9.9\pm0.7)\times10^{8}$ & \citet{2021ApJ...921...97J}\\
                &       & CO(11--10)& 0.64$\pm$0.05 & $(4.9\pm0.4)\times10^{8}$ & this work\\
                &       & CO(12--11)& 0.45$\pm$0.07 & $(3.8\pm0.6)\times10^{8}$ & this work\\
SPT0311$-$58E$^{\dagger}$   & 6.923 & CO(6--5)  & 0.18$\pm$0.03 & $(7.6\pm1.3)\times10^{7}$ & \citet{2021ApJ...921...97J}\\
                &       & CO(7--6)  & 0.17$\pm$0.04 & $(8.4\pm2.0)\times10^{7}$ & \citet{2021ApJ...921...97J}\\
                &       & CO(10--9) & 0.14$\pm$0.04 & $(9.8\pm2.8)\times10^{7}$ & \citet{2021ApJ...921...97J}\\
                &       & CO(11--10)& 0.16$\pm$0.03 & $(1.2\pm0.2)\times10^{8}$ & this work\\
                &       & CO(12--11)& 0.15$\pm$0.04 & $(1.3\pm0.3)\times10^{8}$ & this work\\
\enddata
\tablecomments{
$^{\dagger}$ Gravitationally lensed source. Fluxes and luminosities are apparent values.
}
\end{deluxetable*}

\begin{deluxetable*}{lclccc}
\tablewidth{0pt}
\tablecaption{CO line fluxes and luminosities for the quasar sample analyzed in this work.\label{tab:co_flux_lum_quasar_thiswork}}
\tablehead{
\colhead{Target} & \colhead{$z$} & \colhead{CO transition} & \colhead{Flux} & \colhead{Luminosity} & \colhead{Reference}\\
 &  &  & \colhead{[Jy km s$^{-1}$]} & \colhead{[$L_\odot$]} & }
\startdata
BR1202$-$0725 & 4.687 & CO(1--0) & 0.120$\pm$0.010 & $(4.7\pm0.4)\times10^{6}$ & \citet{2006ApJ...650..604R} \\
 &  & CO(2--1) & 0.23$\pm$0.04 & $(1.8\pm0.3)\times10^{7}$ & \citet{2002AJ....123.1838C} \\
 &  & CO(5--4) & 1.6$\pm$0.2 & $(3.2\pm0.4)\times10^{8}$ & \citet{2012AA...545A..57S} \\
 &  & CO(7--6) & 2.3$\pm$0.2 & $(6.3\pm0.6)\times10^{8}$ & \citet{2012AA...545A..57S}\\
 &  & CO(10--9)  & 1.16$\pm$0.06 & $(4.6\pm0.2)\times10^{8}$ & this work\\
 &  & CO(12--11) & 1.03$\pm$0.05 & $(4.9\pm0.2)\times10^{8}$ & \citet{2021ApJ...913...41L} \\
 &  & CO(13--12) & 1.04$\pm$0.15 & $(5.3\pm0.8)\times10^{8}$ & this work\\
 &  & CO(14--13) & 0.68$\pm$0.10 & $(3.7\pm0.6)\times10^{8}$ & this work\\
J2310+1855 & 6.003 & CO(2--1) & 0.18$\pm$0.02 & $(2.1\pm0.2)\times10^{7}$ & \citet{2019ApJ...876...99S}\\
 &  & CO(5--4) & 0.89$\pm$0.09 & $(2.5\pm0.3)\times10^{8}$ & \citet{2020ApJ...889..162L}\\
 &  & CO(6--5) & 1.12$\pm$0.06 & $(3.8\pm0.2)\times10^{8}$ & \citet{2020ApJ...889..162L}\\
 &  & CO(8--7) & 1.53$\pm$0.05 & $(7.0\pm0.2)\times10^{8}$ & \citet{2020ApJ...889..162L}\\
 &  & CO(9--8) & 1.31$\pm$0.06 & $(6.7\pm0.3)\times10^{8}$ & \citet{2020ApJ...889..162L}\\
 &  & CO(10--9) & 1.04$\pm$0.17 & $(5.9\pm1.0)\times10^{8}$ & \citet{2020ApJ...889..162L}\\
 &  & CO(12--11) & 0.78$\pm$0.13 & $(5.3\pm0.9)\times10^{8}$ & \citet{2020ApJ...889..162L}\\
 &  & CO(13--12) & 0.66$\pm$0.12 & $(4.9\pm0.9)\times10^{8}$ & this work\\
 &  & CO(14--13) & 0.41$\pm$0.11 & $(3.3\pm0.9)\times10^{8}$ & this work\\
J2219+0102 & 6.149 & CO(6--5) & 0.18$\pm$0.03 & $(6.4\pm1.1)\times10^{7}$ & \citet{2022AA...662A..60D}\\
 &  & CO(7--6) & 0.28$\pm$0.04 & $(1.2\pm0.2)\times10^{8}$ & this work\\
 &  & CO(9--8) & 0.11$\pm$0.02 & $(5.9\pm1.1)\times10^{7}$ & this work\\
J0100+2802 & 6.328 & CO(2--1) & 0.38$\pm$0.013 & $(4.7\pm0.2)\times10^{7}$ & \citet{2016ApJ...830...53W}\\
 &  & CO(6--5) & 0.26$\pm$0.05 & $(9.6\pm1.9)\times10^{7}$ & \citet{2019ApJ...880....2W}\\
 &  & CO(7--6) & 0.22$\pm$0.06 & $(9.5\pm2.6)\times10^{7}$ & \citet{2019ApJ...880....2W}\\
 &  & CO(10--9) & 0.25$\pm$0.05 & $(1.5\pm0.3)\times10^{8}$ & \citet{2019ApJ...880....2W}\\
 &  & CO(11--10) & 0.28$\pm$0.04 & $(1.9\pm0.3)\times10^{8}$ & \citet{2019ApJ...880....2W}\\
 &  & CO(12--11) & 0.41$\pm$0.11 & $(3.0\pm0.8)\times10^{8}$ & this work\\
 &  & CO(16--15) & 0.30$\pm$0.04 & $(3.0\pm0.4)\times10^{8}$ & this work\\
 &  & CO(17--16) & 0.27$\pm$0.04 & $(2.8\pm0.4)\times10^{8}$ & this work\\
PJ036+03 & 6.541 & CO(6--5) & 0.35$\pm$0.03 & $(1.4\pm0.1)\times10^{8}$ & \citet{2022AA...662A..60D}\\
 &  & CO(7--6) & 0.40$\pm$0.04 & $(1.8\pm0.2)\times10^{8}$ & \citet{2022AA...662A..60D}\\
 &  & CO(9--8) & 0.23$\pm$0.04 & $(1.3\pm0.2)\times10^{8}$ & this work\\
 &  & CO(10--9) & 0.28$\pm$0.05 & $(1.8\pm0.3)\times10^{8}$ & this work\\
J2348$-$3054 & 6.903 & CO(6--5) & 0.28$\pm$0.05 & $(1.2\pm0.2)\times10^{8}$ & \citet{2017ApJ...845..154V}\\
 &  & CO(7--6) & 0.26$\pm$0.06 & $(1.3\pm0.3)\times10^{8}$ & \citet{2017ApJ...845..154V}\\
 &  & CO(9--8) & 0.26$\pm$0.06 & $(1.6\pm0.4)\times10^{8}$ & this work\\
 &  & CO(10--9) & 0.15$\pm$0.03 & $(1.0\pm0.2)\times10^{8}$ & this work\\
\enddata
\end{deluxetable*}

\subsection{Literature compilation}\label{sec:literature_catalog}

For the SMG and quasar samples, high-$J$ CO detections ($J_{\rm up}\ge 9$) have been reported in the literature for a subset of well-studied sources.
To incorporate these published measurements into the CO SLED analysis presented in Section~\ref{sec:results}, we compiled a supplementary catalog of literature CO line fluxes for SMGs and quasars (Appendix~\ref{app:co_flux_literature}).
For each entry, we adopt the velocity-integrated line fluxes and redshifts reported in the original references.
Although motivated by the availability of high-$J$ measurements, we include lower-$J$ transitions where available for the same sources to extend the SLED coverage.
These literature fluxes are measured with methods that differ between studies, which adds some heterogeneity to the sample.
In addition, our own fluxes contain on average a fraction of $0.93 \pm 0.02$ of the total Gaussian flux, so where the SLED of a source combines our values with literature total fluxes, the affected ratios carry a systematic offset of about 7\%.
This offset is within the literature flux uncertainties.
The CO line fluxes and luminosities from this work and from the literature, listed in Tables 3 to 7, are also provided together as a single machine-readable catalog.

\section{Results}\label{sec:results}

We construct CO SLEDs for the Hot DOG, SMG, and quasar samples by combining the velocity-integrated CO line fluxes measured in this work with those compiled from the literature, and converting them to line luminosities.
Throughout this paper, we assume a flat $\Lambda$CDM cosmology with $H_{0}=70~{\rm km~s^{-1}~Mpc^{-1}}$ and $\Omega_{\rm m}=0.3$.
The CO line luminosity in solar units is computed as
\begin{equation}
L_{\rm CO}\,[L_{\odot}] = 1.04\times10^{-3}\, (S\Delta v)\,
\nu_{\rm rest}\,
\frac{D_{\rm L}^{2}}{(1+z)},
\label{eq:lco}
\end{equation}
where $S\Delta v$ is the integrated line flux in ${\rm Jy\,km\,s^{-1}}$, $\nu_{\rm rest}$ is the rest frequency in GHz, and $D_{\rm L}$ is the luminosity distance in Mpc.

Because the set of observed CO transitions varies from source to source, we construct continuous CO SLED curves by interpolating between the measured points in $\log L_{\rm CO}$--$J_{\rm up}$ space.
We adopt a shape-preserving Piecewise Cubic Hermite Interpolating Polynomial (PCHIP), which passes through all measured line luminosities by construction and reduces spurious oscillations relative to standard cubic spline interpolation.
No extrapolation is performed beyond the $J_{\rm up}$ range spanned by the measured transitions for each source.

\begin{figure*}
  \centering
  \includegraphics[scale=1.0]{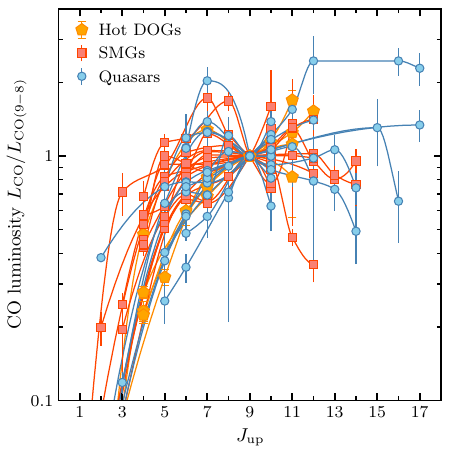}\hspace{3mm}
  \includegraphics[scale=1.0]{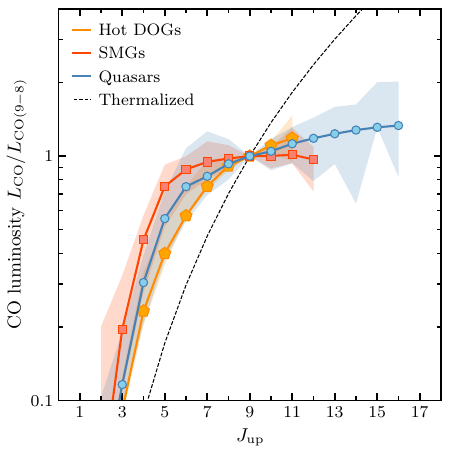}
  \caption{
(Left) Normalized CO SLEDs of individual high-redshift sources: Hot DOGs (pentagons), SMGs (squares), and quasars (circles).
Curves show the PCHIP interpolation in $\log L_{\rm CO}$--$J_{\rm up}$ space, which passes through all measured points.
(Right) Class-median normalized CO SLEDs for the high-redshift sample.
Solid lines show the class median SLED, and shaded regions indicate the 16th--84th percentile range across sources at each $J_{\rm up}$ (Table \ref{tab:cosled_norm_band}).
For a given source, transitions that are not directly measured are evaluated from the interpolated SLED.
Only bins with at least three contributing sources are shown.
The black dashed line shows a thermalized CO SLED.
  }
  \label{fig:COSLED_individual+classband}
\end{figure*}

To compare SLED shapes independently of the overall luminosity scale, we normalize each source by its CO(9--8) luminosity.
If CO(9--8) is not directly measured, we evaluate $L_{\rm CO(9-8)}$ from the interpolated SLED, provided that $J_{\rm up}=9$ falls within the $J_{\rm up}$ range covered by the measurements.

Figure~\ref{fig:COSLED_individual+classband} presents the normalized CO SLEDs for individual sources in the Hot DOG, SMG, and quasar samples.
The SLED shapes reveal substantial source-to-source diversity within each class.
Most systems rise rapidly from low to mid-$J$ (typically $J_{\rm up}\sim 3$--5 to $J_{\rm up}\sim 7$--9), and only a small fraction of sources in any class exhibit a clear SLED peak at $J_{\rm up}<9$.
At higher $J$, the SLEDs diverge, with some sources showing a turnover and decline by $J_{\rm up}\sim 10$--12, whereas others remain approximately flat or continue to rise toward the highest-$J$ transitions observed.
In several cases with coverage extending to $J_{\rm up}=11$--12, the SLED shows no evidence for a turnover and remains flat or continues to rise at the highest observed transition, suggesting that the peak likely lies at $J_{\rm up}>12$ and pointing to the presence of a very highly excited molecular gas component.

Given the heterogeneous transition coverage and large intrinsic scatter, class-to-class differences are difficult to discern from individual SLEDs alone.
To characterize the typical excitation shape of each class, we evaluate each normalized, interpolated SLED on an integer $J_{\rm up}$ grid and compute the class median at each $J_{\rm up}$ (Figure~\ref{fig:COSLED_individual+classband}).
Consistent with the behavior seen in individual objects, the median SLED in each class rises steeply from low $J$ toward the normalization transition, and none of the class medians exhibits a turnover at $J_{\rm up}<9$.
The three classes broadly converge in the mid- to high-$J$ regime around $J_{\rm up}\sim 8$--11, where the median ratios are close to unity and the inter-class differences are comparable to the intrinsic source-to-source scatter indicated by the 16th--84th percentile bands.

For comparison, we overplot a thermalized CO SLED, corresponding to a brightness temperature that is constant with $J$ and giving $L_{\rm CO}\propto\nu_{\rm rest}^3$, normalized at CO(9--8) as for the observed SLEDs (black dashed line in Figure~\ref{fig:COSLED_individual+classband}).
All three class-median SLEDs rise more slowly with $J$ than this relation. At $J_{\rm up}>9$ they lie below the thermalized curve and flatten or turn over around $J_{\rm up}\sim9$--$11$, which indicates that the high-$J$ lines are sub-thermally excited relative to CO(9--8) and that the gas is not fully thermalized up to these levels. At $J_{\rm up}<9$ the SLEDs lie above the thermalized curve, which indicates higher brightness temperatures in the lower-$J$ lines than in CO(9--8).

Among the three classes, at lower $J$ ($J_{\rm up}\sim 4$--6) the SMG median lies systematically above the Hot DOG and quasar medians, indicating stronger low- and mid-$J$ emission compared to CO(9--8) in the typical SMG.
In contrast, the Hot DOG median shows the steepest rise toward $J_{\rm up}\approx 9$ and the lowest median ratios at $J_{\rm up}\lesssim 7$.
Beyond the normalization point, the SMG median remains approximately flat at the CO(9--8) level out to $J_{\rm up}\sim 12$, while the Hot DOG median shows a modest enhancement above unity around $J_{\rm up}\sim 10$--12 within the quoted class scatter.
The quasar median displays the most extended high-$J$ tail, remaining above unity for $J_{\rm up}\gtrsim 10$ and increasing gradually toward the highest-$J$ bins shown.
We note that the Hot DOG and SMG medians are constrained only up to $J_{\rm up}\sim 12$ because we require at least three sources per bin.
The quasar class provides sufficient coverage to extend the median to $J_{\rm up}=17$, although the percentile band broadens substantially at the highest $J$, reflecting the large diversity in high-$J$ excitation within that class.

\section{Discussion}\label{sec:discussion}
\subsection{Qualitative class differences in excitation}\label{sec:disc_class}

The CO SLEDs of our high-redshift sources exhibit broadly similar excitation shapes across the three classes, but with substantial object-to-object scatter.
To quantify relative excitation with a single empirical metric, we introduce a modified excitation parameter $\alpha'$, inspired by the $\alpha$ parameter of \citet{2015ApJ...801...72R} but adapted to the limited and heterogeneous line coverage of our high-redshift sample.
We define
\begin{equation}
\alpha' \equiv \frac{L_{\rm CO(9-8)} + L_{\rm CO(10-9)}}
{L_{\rm CO(6-5)} + L_{\rm CO(7-6)}}.
\label{eq:alphaprime}
\end{equation}
This ratio compares higher-$J$ emission to that at intermediate $J$ and serves as a single-valued parameter for the degree of excitation.

\begin{figure*}
  \centering
  \includegraphics[scale=1.0]{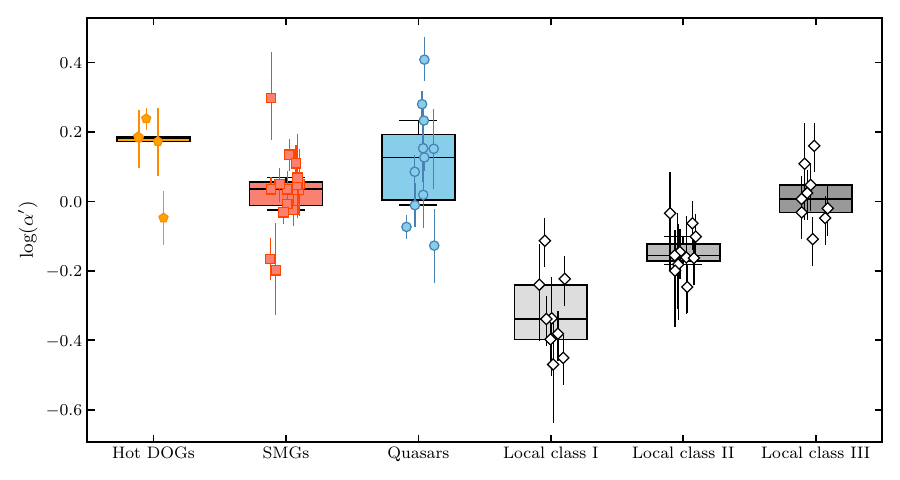}
\caption{
Distribution of the excitation parameter $\alpha'$ for the three high-redshift classes and the three local excitation classes defined by \citet{2015ApJ...801...72R}.
Points indicate per-object median $\alpha'$ values with 16th--84th percentile error bars from Monte Carlo propagation of measurement uncertainties.
Boxes show the interquartile range (25th--75th percentiles) with the median, and whiskers indicate the 16th--84th percentile range.
}
\label{fig:alpha_prime}
\end{figure*}

For each high-redshift source, we compute $\alpha'$ using the PCHIP-interpolated $\log L_{\rm CO}$--$J_{\rm up}$ curve.
When one or more of the four transitions in Equation~(\ref{eq:alphaprime}) is not directly measured, the corresponding luminosity is evaluated from the interpolated SLED, provided that the measured transitions bracket the required $J_{\rm up}=6$--10 interval.
No extrapolation beyond the measured $J_{\rm up}$ coverage is performed, and sources that do not meet this criterion are excluded from the $\alpha'$ analysis.

Uncertainties on $\alpha'$ are estimated via Monte Carlo propagation of the measurement errors.
For each source, we generate $2\times10^{4}$ realizations of the measured line luminosities by drawing $\log L_{\rm CO}$ from Gaussian distributions whose standard deviations are set by the reported line-flux uncertainties.
The interpolated SLED is reconstructed and $\alpha'$ recomputed for each realization.
We adopt the median of the resulting $\alpha'$ distribution as the best estimate and the 16th--84th percentile range as the uncertainty (Table \ref{tab:ir_lum_mu}).

To place the high-redshift measurements in the context of well-studied nearby systems, we also compute $\alpha'$ for the Herschel Comprehensive ULIRG Emission Survey (HerCULES) sample presented by \citet{2015ApJ...801...72R}.
HerCULES is a flux-limited sample of 29 local (U)LIRGs spanning $L_{\rm IR}\sim 10^{11}$--$10^{13}\,L_\odot$, for which \textit{Herschel}/SPIRE provided CO observations from $J_{\rm up}=4$ through 13.
We assign the local galaxies to excitation classes~I, II, and III following the classification of \citet{2015ApJ...801...72R}, defined using their $\alpha$ parameter based on the relative strength of high-$J$ to mid-$J$ CO lines.

Figure~\ref{fig:alpha_prime} shows the distribution of $\alpha'$ for the three high-redshift classes together with the three local classes.
For the high-redshift sample, the Hot DOGs have a median $\alpha' = 1.52$ (16th--84th percentile range $1.28$--$1.61$, $N=5$), the SMGs have $\alpha' = 1.08$ ($0.94$--$1.15$, $N=16$), and the quasars have $\alpha' = 1.34$ ($0.92$--$1.79$, $N=11$).
For the local sample, the class~I, II, and III galaxies have median $\alpha' = 0.46$ ($0.37$--$0.59$, $N=9$), $0.70$ ($0.65$--$0.82$, $N=11$), and $1.02$ ($0.91$--$1.24$, $N=9$), respectively.
These values indicate substantial intra-class scatter within the high-redshift populations and only modest shifts among their class medians, whereas the local classes span a wider range in $\alpha'$.

To compare the excitation parameter $\alpha'$ between the classes, we use the two-sided Mann--Whitney $U$ test. We adopt this test because the samples are small and the $\alpha'$ values are skewed luminosity ratios, so a rank-based test that assumes no particular distribution shape and is robust to outliers is more suitable than a test that assumes a normal distribution, and it directly evaluates whether the $\alpha'$ values of one class are stochastically larger than those of another, which is our question of which class is more highly excited.
We report the Mann--Whitney $p$-value and the rank-biserial correlation as an effect size, where positive values indicate larger $\alpha'$ in the first-named sample.
Motivated by the qualitative impression from the normalized CO SLEDs that Hot DOGs may exhibit more highly excited CO emission than SMGs, we treat the Hot DOG versus SMG comparison as the primary hypothesis test.
The Hot DOGs ($N=5$) show a tendency toward larger $\alpha'$ than the SMGs ($N=16$), with $r_{\rm rb}=0.55$, but the difference is not statistically significant ($p=0.08$).
We therefore regard the apparent offset as suggestive rather than conclusive.
The limited sample sizes and substantial object-to-object scatter likely dominate the uncertainty in assessing modest class-dependent shifts among the high-redshift populations.

We next compare the pooled high-redshift sample ($N=32$) with the three local excitation classes.
We test whether the high-redshift distribution is consistent with each local class individually using two-sided Mann--Whitney tests.
The pooled high-redshift sample is strongly shifted toward larger $\alpha'$ relative to the low-excitation local classes, with high-$z$ versus class~I giving $r_{\rm rb}=0.98$ and $p=9.6\times10^{-6}$, and high-$z$ versus class~II giving $r_{\rm rb}=0.87$ and $p=2.2\times10^{-5}$.
The comparison to the high-excitation local class~III, which comprises extreme systems such as Arp~220, NGC~6240, and Mrk~231 where mechanical heating or AGN-driven X-ray irradiation contributes significantly to the CO excitation, yields a smaller offset, with $r_{\rm rb}=0.34$ and $p=0.13$.
Taken together, these results indicate that the high-redshift sources are inconsistent with the low-excitation local populations and are broadly comparable to the most highly excited nearby galaxies, although any residual difference relative to class~III remains modest and will require larger samples with more uniform CO line coverage to assess definitively.

\subsection{Physical interpretation}\label{sec:disc_phys}

\begin{figure*}
  \centering
  \includegraphics[scale=1.0]{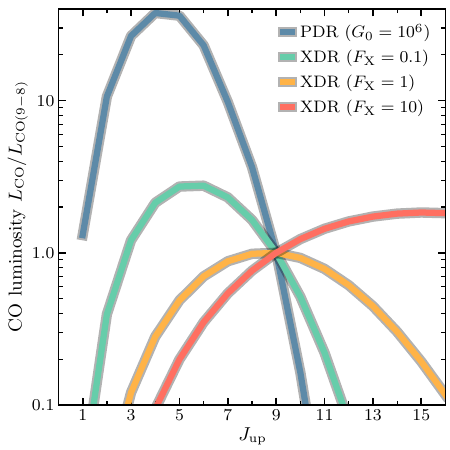}\hspace{3mm}
  \includegraphics[scale=1.0]{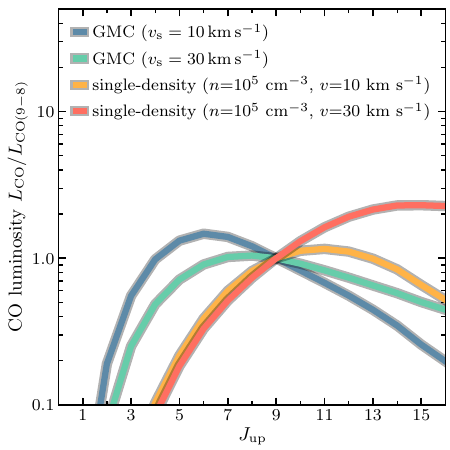}
  \caption{
(Left) Predicted CO spectral line energy distributions (SLEDs) for a single GMC ($10^{4.5}\,M_\odot$) under different excitation conditions.
Shown are a far-ultraviolet--irradiated photodissociation region (PDR; $G_0=10^6$; blue) and X-ray--dominated region (XDR) models with incident X-ray fluxes $F_{\rm X}=0.1$, 1, and 10~erg~s$^{-1}$~cm$^{-2}$ (green, orange, and red, respectively).
The CO line luminosities are expressed as ratios normalized to the CO(9--8) transition, $L_{\rm CO}/L_{\rm CO(9-8)}$.
(Right) Predicted CO SLEDs for C-type shock models \citep{2015AA...578A..63F}.
Single-density models at $\log\,n_{\rm H}/{\rm cm}^{-3} = 5$ are shown for $v_{\rm s}=10$ (orange) and $30~\mathrm{km~s^{-1}}$ (red), along with GMC models at the same velocities (blue and green, respectively; Appendix~\ref{app:gmc_cshock}).
}
  \label{fig:COSLED_XDR_shock}
\end{figure*}

The broad consistency between the high-redshift sources and the local class~III systems motivates a closer examination of which excitation mechanisms can reproduce the observed flat high-$J$ CO SLEDs.
We compare the observed SLED shapes with PDR, XDR, and shock models to assess whether radiative or mechanical heating better accounts for the high-$J$ behavior.

We first consider the PDR/XDR models shown in Figure~\ref{fig:COSLED_XDR_shock}, which are based on physically motivated giant molecular cloud (GMC) models \citep{2019MNRAS.490.4502V,2024MNRAS.527.8727E}.
Even for an extreme far-ultraviolet field ($G_0=10^6$), the PDR model peaks at low--mid $J$ and drops rapidly at $J_{\rm up}\gtrsim 8$.
Such a PDR-dominated SLED would predict low-$J$ lines much brighter than CO(9--8), inconsistent with the relatively suppressed low-$J$/CO(9--8) ratios seen in Figure~\ref{fig:COSLED_individual+classband}.
The observed SLEDs are therefore difficult to reconcile with a scenario in which a single PDR component dominates the CO excitation.
In contrast, the XDR models shift the excitation toward higher $J$ as the incident X-ray flux increases, producing progressively flatter high-$J$ tails.
The high-flux XDR cases ($F_{\rm X}=1$--$10$~erg~s$^{-1}$~cm$^{-2}$) provide a closer match to the observed flat high-$J$ behavior, consistent with the finding that XDR-heated gas can dominate the CO luminosity at $J_{\rm up}\gtrsim 4$ in AGN-host galaxies \citep{2024MNRAS.527.8727E}.
The tentative enhancement of $\alpha'$ in the Hot DOGs relative to the SMGs (Section~\ref{sec:disc_class}) may reflect this picture.
Hot DOGs are thought to harbor luminous but heavily dust-obscured AGN whose X-ray emission is strongly absorbed along the line of sight, making them difficult to identify through conventional X-ray surveys.
Because high-$J$ CO emission originates deep within the molecular medium and is largely unaffected by dust extinction, it may offer an indirect probe of such obscured X-ray sources, complementary to direct X-ray and mid-infrared diagnostics.


The comparison above is qualitative. 
To put it on a quantitative footing, we fit each class-median SLED with a two-component model that combines the PDR and XDR GMC models of \citet{2024MNRAS.527.8727E}, following an approach similar to \citet{2024ApJ...962..119L}.
We fix the PDR field at $\log G_0 = 6$, the extreme far-ultraviolet field used in Figure~\ref{fig:COSLED_XDR_shock}, and vary the incident X-ray flux $\log F_{\rm X}$ together with the fraction $f$ of the CO(9--8) luminosity carried by the XDR component, so that $R(J) = (1-f)\,R_{\rm PDR}(J) + f\,R_{\rm XDR}(J)$.
We fit in log space over all available transitions, using the half-width of the 16th to 84th percentile band as the uncertainty and excluding the CO(9--8) normalization point.
The XDR component carries almost all of the CO(9--8) luminosity in every class, with $f \simeq 0.99$ for the SMGs and quasars and $f=1$ for the Hot DOGs. The best-fit X-ray fluxes are $\log F_{\rm X} = 0.25^{+0.07}_{-0.05}$ for the Hot DOGs, $0.25^{+0.14}_{-0.09}$ for the SMGs, and $0.75^{+0.11}_{-0.08}$ for the quasars, with reduced $\chi^2$ between 0.6 and 1.5.
When $\log G_0$ is instead left free, the SMG and quasar fits drive it to the top of the explored range, while for the Hot DOGs it is unconstrained because the PDR contribution already vanishes.
In all cases the best-fit $\log F_{\rm X}$ and $f$ are essentially the same, which is why we fix $\log G_0 = 6$. Although the PDR share of CO(9--8) is small, the steep PDR SLED makes it the main source of the low-$J$ lines. 
In the SMGs and quasars the PDR supplies most of the emission up to CO(4--3) and about half at CO(5--4), and the XDR takes over above $J_{\rm up}\simeq 6$. The Hot DOGs require no PDR component and are thus the most strongly XDR-dominated class, consistent with their high $\alpha'$. Adding the PDR component improves the fit over a pure XDR model for the SMGs and quasars, while for the Hot DOGs a pure XDR model is already sufficient.


Mechanical heating by shocks provides an alternative route to producing highly excited CO SLEDs, potentially mimicking those produced by intense X-ray irradiation.
Figure~\ref{fig:COSLED_XDR_shock} presents four representative C-type shock models from \citet{2015AA...578A..63F}, which are parameterized by the preshock density, $n_{\rm H}$ (cm$^{-3}$), and the shock velocity, $v_{\rm s}$.
These include single-density models at $\log\,n_{\rm H}/{\rm cm}^{-3}=5$ with $v_{\rm s}=10$ and $30~{\rm km~s^{-1}}$, as well as GMC-integrated models at the same velocities (Section~\ref{app:gmc_cshock}).
The single-density models remain strongly excited and can sustain bright emission at $J_{\rm up}\gtrsim 9$.
However, a medium uniformly composed of dense ($\log\,n_{\rm H}/{\rm cm}^{-3}=5$) gas is physically unrealistic on galaxy scales.
When the emission is integrated over a more plausible clump density distribution, the CO SLED becomes dominated by the more numerous intermediate-density clumps at $\log\,n_{\rm H}/{\rm cm}^{-3}\sim 3$--4 (Appendix~\ref{app:gmc_cshock}).
This naturally boosts the low-$J$ lines and produces a comparatively weaker high-$J$ tail, and even the GMC model with $v_{\rm s}=30~{\rm km~s^{-1}}$ struggles to maintain strong excitation beyond $J_{\rm up}\gtrsim 9$.
Taken together, these comparisons imply that reproducing a strongly excited CO SLED with shocks alone requires additional conditions, such as shocks that dissipate preferentially in the densest phase or an unusually large dense-gas fraction.

A spatially unresolved CO SLED alone does not uniquely identify the dominant heating mechanism, since both intense X-ray irradiation and dense, shock-heated gas can produce elevated high-$J$ emission.
Discriminating between XDR- and shock-dominated excitation thus requires additional observables that probe where the high-$J$ emission arises and how it relates to dust heating and gas dynamics.
Spatially resolved multi-transition CO mapping would directly test whether the high-$J$ emission is centrally concentrated, as expected for compact XDR-dominated regions, or spatially extended and correlated with disturbed kinematics, as expected for merger-driven turbulence, outflows, or large-scale shocks \citep{2025NatAs...9..720T}.

\begin{figure*}
  \centering
  \includegraphics[scale=1.0]{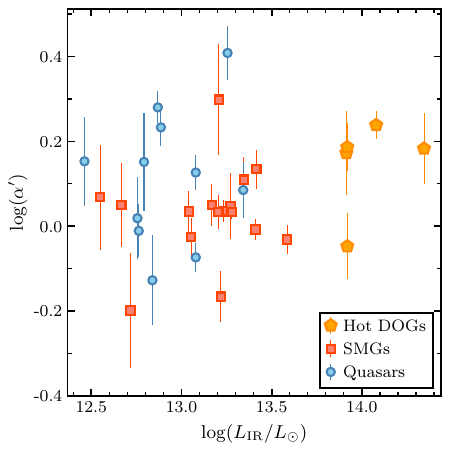}
  \includegraphics[scale=1.0]{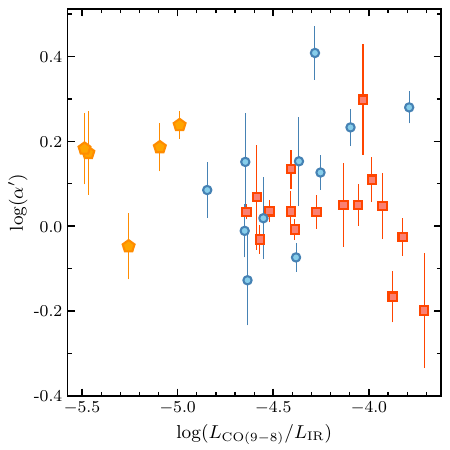}
  \caption{
(Left) Relation between the excitation parameter and intrinsic infrared luminosity (corrected for gravitational lensing).
(Right) Relation between the excitation parameter and CO(9--8)-to-infrared luminosity ratio.
}
  \label{fig:LIR_alpha_prime}
\end{figure*}

\subsection{Excitation, infrared luminosity, and the CO-to-infrared ratio}\label{sec:disc_lir_alpha}

As discussed in Section~\ref{sec:disc_phys}, the shape of a spatially unresolved CO SLED alone does not uniquely discriminate between XDR heating and shock heating.
A complementary diagnostic is the infrared luminosity, $L_{\rm IR}$, which traces the radiative power absorbed and re-emitted by dust.
In a simple XDR picture, an energetically dominant AGN can both excite molecular gas and heat dust efficiently, potentially producing high-$J$ CO emission together with large $L_{\rm IR}$.
Conversely, mechanical heating can enhance high-$J$ CO emission by depositing kinetic energy directly into the gas while producing a much weaker response in the far-infrared continuum \citep[e.g.,][]{2013ApJ...762L..16M}.
However, a comparison of $\alpha^\prime$ with $L_{\rm IR}$ reveals no clear trend across the full sample (Figure~\ref{fig:LIR_alpha_prime} and Table~\ref{tab:ir_lum_mu}).

To obtain a metric that more directly reflects how energy is partitioned between molecular gas cooling and dust emission, we examine the CO(9--8)-to-infrared luminosity ratio, $L_{\rm CO(9-8)}/L_{\rm IR}$, and its relation to $\alpha^\prime$ (Figure~\ref{fig:LIR_alpha_prime}).
A Spearman test on the full sample yields $\rho = -0.10$ with $p = 0.60$ ($N=32$), consistent with no global monotonic relation between $\alpha^\prime$ and $L_{\rm CO(9-8)}/L_{\rm IR}$.
Restricting the comparison to the dusty galaxy populations (Hot~DOGs and SMGs) yields a more suggestive, though still not formally significant, anti-correlation ($\rho = -0.39$, $p = 0.08$, $N=21$).
This behavior manifests primarily as a systematic offset between the two populations.
Hot~DOGs tend to occupy the region of large $\alpha^\prime$ but small $L_{\rm CO(9-8)}/L_{\rm IR}$, whereas SMGs extend to higher $L_{\rm CO(9-8)}/L_{\rm IR}$ at more moderate $\alpha^\prime$.
The quasars populate a similar range in $L_{\rm CO(9-8)}/L_{\rm IR}$ as the SMGs but show substantial scatter in $\alpha^\prime$, diluting any population-wide trend when all three classes are combined.

The tentative anti-correlation in the dusty galaxy subset implies that increasing excitation (larger $\alpha^\prime$) does not coincide with enhanced high-$J$ CO output per unit infrared luminosity.
This is difficult to reconcile with a scenario in which mechanical heating universally boosts high-$J$ CO emission relative to the dust continuum, because such a mechanism would predict $L_{\rm CO(9-8)}/L_{\rm IR}$ to rise with $\alpha^\prime$.
A more natural explanation is that the infrared luminosity in Hot~DOGs is dominated by compact, AGN-heated dust whose output reflects the bolometric power of the nucleus.
In this picture, XDR heating of the surrounding molecular medium can elevate $\alpha^\prime$ by preferentially enhancing the high-$J$ portion of the CO SLED, but $L_{\rm IR}$ rises even more steeply because it is driven by the total AGN luminosity reprocessed by dust.
The net result is that $L_{\rm CO(9-8)}/L_{\rm IR}$ decreases even as $\alpha^\prime$ grows, producing the observed anti-correlation.
The absence of a clear trend among quasars may reflect a similar decoupling between AGN-dominated infrared output and molecular-gas excitation, compounded by source-to-source diversity in heating mechanisms (e.g., mergers, outflows, or localized shocks) that introduces scatter and obscures any simple monotonic relation.
Establishing whether the offset between Hot~DOGs and SMGs reflects a genuine physical sequence will require larger samples with more uniform coverage of $J_{\rm up}\sim6$--10 and infrared SED decompositions that separate AGN-heated and star-formation-heated dust components.

\section{Summary}\label{sec:summary}

We have presented a catalog of high-$J$ CO measurements for galaxies at $z\gtrsim 3$ based on the ALMA Science Archive, enabling comparative studies of molecular-gas excitation across high-redshift galaxy populations.
Our main results are summarized below.

\begin{enumerate}

\item We searched the ALMA Science Archive to identify CO spectroscopy suitable for high-$J$ CO measurements at $z>3$.
We selected sources spanning three broad classes (Hot DOGs, SMGs, and optically selected quasars) and performed a homogeneous reduction of the archival CO datasets.
We fitted the CO spectra with Gaussian profiles to define velocity integration ranges and to correct for blending with nearby transitions.
We report new velocity-integrated CO line fluxes and derived luminosities.

\item We constructed continuous CO SLEDs by interpolating the measured $\log L_{\rm CO}$ values as a function of $J_{\rm up}$, normalized to CO(9--8).
The individual normalized SLEDs show substantial source-to-source diversity within each class.
Only a small fraction of sources show evidence for a peak at $J_{\rm up}<9$, and in several cases the absence of a turnover by $J_{\rm up}\simeq 11$--12 suggests a SLED peak at even higher $J$, consistent with the presence of very highly excited molecular gas.

\item Class-median normalized SLEDs show broadly similar excitation shapes across the three classes, but with systematic differences.
The SMG median exhibits relatively stronger low- and mid-$J$ emission compared to CO(9--8), while the Hot DOG median rises more steeply toward $J_{\rm up}\simeq 9$ and shows modest enhancement at $J_{\rm up}\sim 10$--12.
Using the empirical excitation parameter $\alpha'$, defined as the ratio of $J_{\rm up}=9$--10 to $J_{\rm up}=6$--7 emission, the Hot DOGs show a tendency toward larger $\alpha'$ than the SMGs, although the difference is not statistically significant for the present sample.

\item Compared to the local (U)LIRG sample, the high-redshift sources are strongly shifted toward higher excitation relative to low-excitation local systems and are broadly comparable to class~III galaxies such as Arp~220, NGC~6240, and Mrk~231, where mechanical heating or AGN-driven X-ray irradiation contributes significantly to the CO excitation.

\item Comparison with simple excitation models indicates that a single PDR component is difficult to reconcile with the observed flat high-$J$ behavior, while XDR-like heating or dense, shock-heated gas can reproduce more extended high-$J$ emission.
If shocks dominate, maintaining a flat high-$J$ SLED requires shocks that dissipate preferentially in the densest phase or an unusually large dense-gas fraction.

\item Hot~DOGs tend to exhibit smaller $L_{\rm CO(9-8)}/L_{\rm IR}$ than SMGs despite their higher $\alpha^\prime$, suggesting a tentative anti-correlation between excitation and high-$J$ CO output per unit infrared luminosity.
If confirmed with larger samples, this pattern would favor a picture in which the enhanced excitation in Hot~DOGs is driven primarily by XDR heating from an obscured AGN rather than by mechanical heating from shocks, because shocks would be expected to boost $L_{\rm CO(9-8)}/L_{\rm IR}$ rather than suppress it.

\end{enumerate}

With uniform data reduction and a consistently defined excitation parameter, this catalog provides a reference set against which future systematic surveys targeting statistically complete samples can be compared.
The source-integrated high-$J$ CO fluxes and SLED shapes compiled here also provide a practical baseline for prioritizing targets and estimating the sensitivity requirements for high-angular-resolution follow-up observations aimed at resolving the spatial origin of extreme excitation (Paper~II; \citealt{2026arXiv260301352T}).




\begin{acknowledgments}
The author thanks the anonymous referee for the careful reading of the manuscript and for the constructive comments that improved this paper. 
This paper makes use of the following ALMA data: 
ADS/JAO.ALMA\#2012.1.00844.S,
ADS/JAO.ALMA\#2015.1.00504.S,
ADS/JAO.ALMA\#2015.1.00928.S,
ADS/JAO.ALMA\#2015.1.01042.S,
ADS/JAO.ALMA\#2016.1.00330.S,
ADS/JAO.ALMA\#2017.1.00963.S,
ADS/JAO.ALMA\#2017.1.01516.S,
ADS/JAO.ALMA\#2017.A.00032.S,
ADS/JAO.ALMA\#2018.1.00081.S,
ADS/JAO.ALMA\#2018.1.00861.S,
ADS/JAO.ALMA\#2018.1.00966.S,
ADS/JAO.ALMA\#2019.1.00080.S,
ADS/JAO.ALMA\#2019.1.00147.S,
ADS/JAO.ALMA\#2019.1.00466.S,
ADS/JAO.ALMA\#2019.1.00533.S,
ADS/JAO.ALMA\#2021.1.00168.S,
ADS/JAO.ALMA\#2021.1.01015.S,
ADS/JAO.ALMA\#2021.1.01313.S,
ADS/JAO.ALMA\#2022.1.00346.S,
ADS/JAO.ALMA\#2022.1.00353.S,
ADS/JAO.ALMA\#2023.1.00653.S,
ADS/JAO.ALMA\#2023.1.00852.S,
ADS/JAO.ALMA\#2023.1.01281.S,
ADS/JAO.ALMA\#2023.1.01532.S.
ALMA is a partnership of ESO (representing its member states), NSF (USA) and NINS (Japan), together with NRC (Canada), NSTC and ASIAA (Taiwan), and KASI (Republic of Korea), in cooperation with the Republic of Chile. The Joint ALMA Observatory is operated by ESO, AUI/NRAO and NAOJ.
We are grateful to the staff of the East Asian ALMA Regional Center for their support in providing calibrated measurement sets.
The author thanks Sougo Hasegawa for carrying out the initial ALMA data analysis of SPT0311$-$58 as part of his undergraduate research project.
The author used ChatGPT (OpenAI) and Claude (Anthropic) for assistance with language editing and code refinement. The author takes full responsibility for the content of this manuscript.
K.T. was supported by the ALMA Japan Research Grant of NAOJ ALMA Project, NAOJ-ALMA-387.
K.T. acknowledges support from JSPS KAKENHI Grant Number JP 23K03466.
\end{acknowledgments}

%
\facilities{ALMA}


\software{CASA \citep{2022PASP..134k4501C}, galaxySLED \citep{2024MNRAS.527.8727E}, SciPy \citep{2020SciPy-NMeth}}



\appendix

\section{Multi-transition CO Spectra}\label{app:spectra}

\begin{figure*}
  \centering
  \includegraphics[scale=1.0]{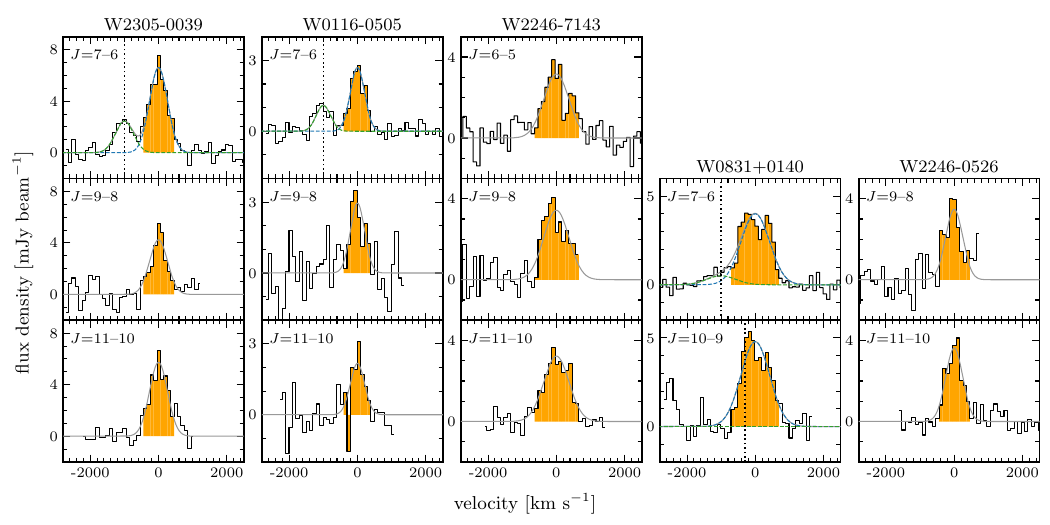}
\caption{
Multi-transition CO spectra for five sources in our Hot DOG sample.
In every panel the black histogram is the observed spectrum and the gray solid curve is the total composite model. 
In the panels where a nearby line is blended with the CO line, the blue dashed and green dashed curves show the CO component and the contaminant, which is [CI](2--1) near CO(7--6) and H$_2$O near CO(10--9), and the vertical dotted line marks the expected position of the contaminant.
In the other panels the CO line is isolated and only the total model is drawn. 
Shaded channels indicate the velocity range adopted for the integrated-flux measurements.
}
\label{fig:spectra_HotDOG}
\end{figure*}

\begin{figure*}[t]
  \centering
  \includegraphics[scale=1.0]{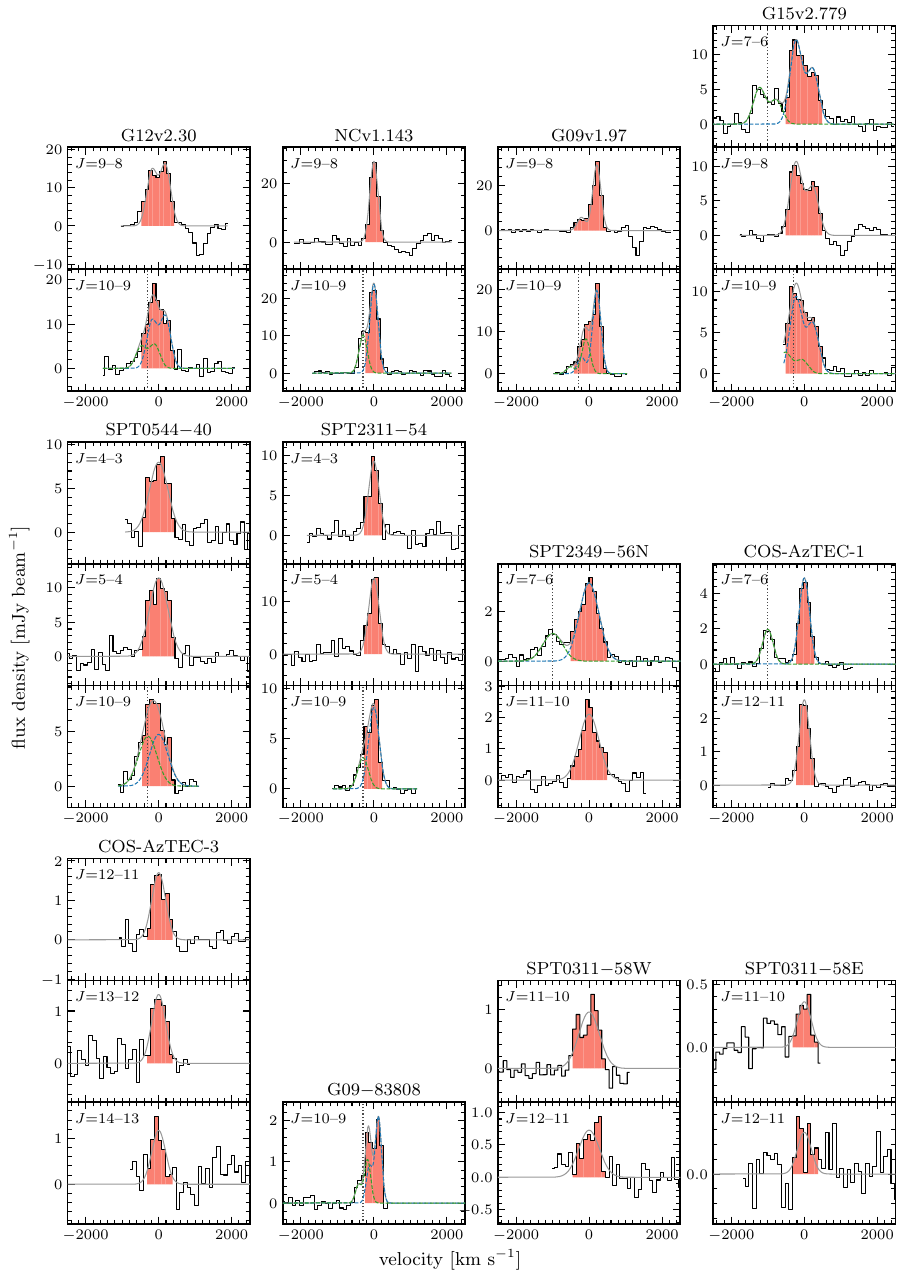}
\caption{
Multi-transition CO spectra for 12 sources in our SMG sample.
}
\label{fig:spectra_SMG}
\end{figure*}

\begin{figure*}
  \centering
  \includegraphics[scale=1.0]{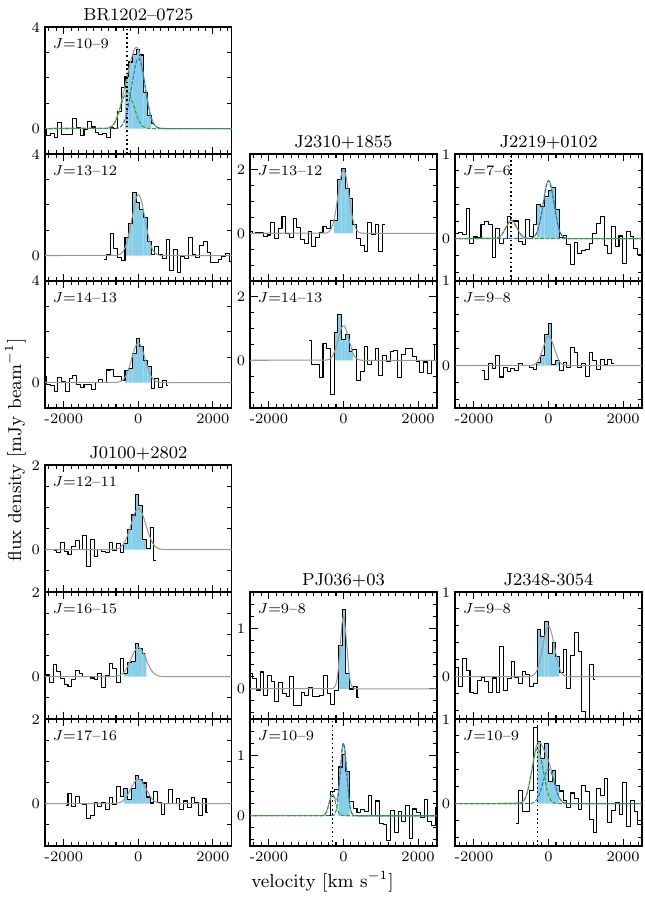}
\caption{
Multi-transition CO spectra for six sources in our quasar sample.
}
\label{fig:spectra_quasar}
\end{figure*}

Figures~\ref{fig:spectra_HotDOG}, \ref{fig:spectra_SMG}, and \ref{fig:spectra_quasar} present the observed multi-transition CO spectra for individual sources in the Hot DOG, SMG, and quasar samples, respectively.

\section{Literature CO Line Measurements}\label{app:co_flux_literature}

Tables~\ref{tab:co_flux_lum_SMG_literature} and~\ref{tab:co_flux_lum_quasar_literature} compile previously published CO line measurements for the SMG and quasar samples.
Each table lists the velocity-integrated line fluxes and redshifts adopted from the cited references, together with line luminosities computed using Equation~(\ref{eq:lco}) and the cosmology assumed throughout this paper.


\begin{deluxetable*}{lclccc}
\tablewidth{0pt}
\tablecaption{Literature CO line fluxes and luminosities for SMGs compiled from the literature.\label{tab:co_flux_lum_SMG_literature}}
\tablehead{
\colhead{Target} & \colhead{$z$} & \colhead{CO transition} & \colhead{Flux} & \colhead{Luminosity} & \colhead{Reference}\\
 &  &  & \colhead{[Jy km s$^{-1}$]} & \colhead{[$L_\odot$]} & }
\startdata
SDP-81  & 3.042 & CO(1--0)  & 1.26$\pm$0.20  & $(2.5\pm0.4)\times10^{7}$ & \citet{2011MNRAS.415.3473V}\\
        &       & CO(3--2)  & 11.8$\pm$2.3  & $(7.0\pm1.4)\times10^{8}$ & \citet{2020MNRAS.494.5542R}\\
        &       & CO(5--4)  & 9.1$\pm$1.0  & $(9.0\pm1.0)\times10^{8}$ & \citet{2015MNRAS.453L..26R}\\
        &       & CO(7--6)  & 12$\pm$4  & $(1.7\pm0.6)\times10^{9}$ & \citet{2012ApJ...757..135L}\\
        &       & CO(8--7)  & 7.6$\pm$0.3  & $(1.2\pm0.05)\times10^{9}$ & \citet{2015MNRAS.453L..26R}\\
        &       & CO(10--9) & 4.5$\pm$0.6  & $(8.9\pm1.2)\times10^{8}$ & \citet{2020MNRAS.494.5542R}\\
ADFS-27 & 5.655 & CO(2--1)  & 0.716$\pm$0.087  & $(7.5\pm0.9)\times10^{7}$ & \citet{2021ApJ...907...62R}\\
        &       & CO(5--4)  & 2.68$\pm$0.20  & $(7.0\pm0.5)\times10^{8}$ & \citet{2021ApJ...907...62R}\\
        &       & CO(6--5)  & 2.82$\pm$0.34  & $(8.8\pm1.1)\times10^{8}$ & \citet{2021ApJ...907...62R}\\
        &       & CO(8--7)  & 2.24$\pm$0.11  & $(9.4\pm0.5)\times10^{8}$ & \citet{2021ApJ...907...62R}\\
        &       & CO(9--8)  & 1.80$\pm$0.04  & $(8.5\pm0.2)\times10^{8}$ & \citet{2021ApJ...907...62R}\\
        &       & CO(10--9) & 2.18$\pm$0.05  & $(1.1\pm0.03)\times10^{9}$ & \citet{2021ApJ...907...62R}\\
SPT0346--52$^{\dagger}$ & 5.656 & CO(2--1)  & 2.15$\pm$0.15  & $(2.2\pm0.2)\times10^{8}$ & \citet{2015ApJ...811..124S}\\
 &  & CO(6--5)  & 11.56$\pm$0.53 & $(3.6\pm0.2)\times10^{9}$ & \citet{2019AA...628A..23A}\\
 &  & CO(8--7)  & 10.30$\pm$0.31 & $(4.3\pm0.1)\times10^{9}$ & \citet{2019AA...628A..23A}\\
 &  & CO(9--8)  & 11.10$\pm$0.96 & $(5.2\pm0.5)\times10^{9}$ & \citet{2019AA...628A..23A}\\
HFLS3 & 6.337 & CO(1--0)  & 0.074$\pm$0.024 & $(4.6\pm1.5)\times10^{6}$ & \citet{2013Natur.496..329R}\\
      &       & CO(2--1)  & 0.32$\pm$0.03   & $(4.0\pm0.4)\times10^{7}$ & \citet{2013Natur.496..329R}\\
      &       & CO(3--2)  & 0.72$\pm$0.09   & $(1.3\pm0.2)\times10^{8}$ & \citet{2013Natur.496..329R}\\
      &       & CO(6--5)  & 2.74$\pm$0.68   & $(1.0\pm0.3)\times10^{9}$ & \citet{2013Natur.496..329R}\\
      &       & CO(7--6)  & 2.22$\pm$0.25   & $(9.6\pm1.1)\times10^{8}$ & \citet{2013Natur.496..329R}\\
      &       & CO(9--8)  & 2.77$\pm$0.45   & $(1.5\pm0.3)\times10^{9}$ & \citet{2013Natur.496..329R}\\
      &       & CO(10--9) & 3.91$\pm$1.59   & $(2.4\pm1.0)\times10^{9}$ & \citet{2013Natur.496..329R}\\
\enddata
\tablecomments{
$^{\dagger}$ Gravitationally lensed source. Fluxes and luminosities are apparent values.
}
\end{deluxetable*}

\begin{deluxetable*}{lclccc}
\tablewidth{0pt}
\tablecaption{Literature CO line fluxes and luminosities for quasars compiled from the literature.\label{tab:co_flux_lum_quasar_literature}}
\tablehead{
\colhead{Target} & \colhead{$z$} & \colhead{CO transition} & \colhead{Flux} & \colhead{Luminosity} & \colhead{Reference}\\
 &  &  & \colhead{[Jy km s$^{-1}$]} & \colhead{[$L_\odot$]} & }
\startdata
APM~08279+5255$^{\dagger}$ & 3.91 & CO(8--7)  & 12.1$\pm$8.3 & $(2.9\pm2.0)\times10^{9}$ & \citet{2011ApJ...741L..37B}\\
 &  & CO(9--8)  & 16.0$\pm$3.8 & $(4.3\pm1.0)\times10^{9}$ & \citet{2011ApJ...741L..37B}\\
 &  & CO(10--9) & 15.2$\pm$7.2 & $(4.5\pm2.1)\times10^{9}$ & \citet{2011ApJ...741L..37B}\\
 &  & CO(11--10)& 14.4$\pm$4.5 & $(4.7\pm1.5)\times10^{9}$ & \citet{2011ApJ...741L..37B}\\
 &  & CO(12--11)& 9.5$\pm$5.0  & $(3.4\pm1.8)\times10^{9}$ & \citet{2011ApJ...741L..37B}\\
J2322+1944$^{\dagger}$ & 4.118 & CO(1--0) & 0.155$\pm$0.013 & $(5.0\pm0.4)\times10^{6}$ & \citet{2006ApJ...650..604R}\\
 &  & CO(9--8) & 3.54$\pm$0.09 & $(1.00\pm0.03)\times10^{9}$ & \citet{2023AA...674L...5B}\\
BRI0952$-$0952$^{\dagger}$ & 4.432 & CO(5--4)   & 1.4$\pm$0.30 & $(2.5\pm0.5)\times10^{8}$ & \citet{2024AA...684A..56K}\\
              &       & CO(7--6)   & 1.7$\pm$0.18 & $(4.3\pm0.5)\times10^{8}$ & \citet{2024AA...684A..56K}\\
              &       & CO(12--11) & 2.0$\pm$0.18 & $(8.7\pm0.8)\times10^{8}$ & \citet{2024AA...684A..56K}\\
J0129$-$0035 & 5.779 & CO(2--1) & 0.036$\pm$0.005 & $(3.9\pm0.5)\times10^{6}$ & \citet{2019ApJ...876...99S}\\
 &  & CO(6--5) & 0.18$\pm$0.03 & $(5.8\pm1.0)\times10^{7}$ & \citet{TripodiPhD}\\
 &  & CO(8--7) & 0.21$\pm$0.09 & $(9.1\pm3.9)\times10^{7}$ & \citet{2024ApJ...977..190X}\\
 &  & CO(9--8) & 0.21$\pm$0.05 & $(1.0\pm0.2)\times10^{8}$ & \citet{2024ApJ...977..190X}\\
P215$-$16 & 5.783 & CO(5--4) & 0.90$\pm$0.17 & $(2.4\pm0.5)\times10^{8}$ & \citet{2024ApJ...962..119L}\\
 &  & CO(6--5) & 1.01$\pm$0.14 & $(3.3\pm0.5)\times10^{8}$ & \citet{2024ApJ...962..119L}\\
 &  & CO(8--7) & 1.54$\pm$0.18 & $(6.7\pm0.8)\times10^{8}$ & \citet{2024ApJ...962..119L}\\
 &  & CO(9--8) & 1.93$\pm$0.19 & $(9.4\pm0.9)\times10^{8}$ & \citet{2024ApJ...962..119L}\\
 &  & CO(10--9) & 2.00$\pm$0.24 & $(1.1\pm0.1)\times10^{9}$ & \citet{2024ApJ...962..119L}\\
J2054$-$0005 & 6.039 & CO(2--1) & 0.06$\pm$0.01 & $(6.9\pm1.2)\times10^{6}$ & \citet{2019ApJ...876...99S}\\
 &  & CO(6--5) & 0.29$\pm$0.05 & $(1.0\pm0.2)\times10^{8}$ & \citet{2024AA...689A.220T}\\
 &  & CO(7--6) & 0.24$\pm$0.08 & $(9.7\pm3.2)\times10^{7}$ & \citet{2022AA...662A..60D}\\
 &  & CO(8--7) & 0.36$\pm$0.07 & $(1.7\pm0.3)\times10^{8}$ & \citet{2024ApJ...977..190X}\\
 &  & CO(9--8) & 0.27$\pm$0.04 & $(1.4\pm0.2)\times10^{8}$ & \citet{2024ApJ...977..190X}\\
 &  & CO(10--9) & 0.24$\pm$0.07 & $(1.4\pm0.4)\times10^{8}$ & \citet{2024ApJ...977..190X}\\
J1429+5447 & 6.184 & CO(5--4) & 0.39$\pm$0.09 & $(1.2\pm0.3)\times10^{8}$ & \citet{2024ApJ...962..119L}\\
 &  & CO(6--5) & 0.53$\pm$0.13 & $(1.9\pm0.5)\times10^{8}$ & \citet{2024ApJ...962..119L}\\
 &  & CO(7--6) & 0.48$\pm$0.10 & $(2.0\pm0.4)\times10^{8}$ & \citet{2024ApJ...962..119L}\\
 &  & CO(9--8) & 0.29$\pm$0.07 & $(1.6\pm0.4)\times10^{8}$ & \citet{2024ApJ...962..119L}\\
 &  & CO(10--9) & 0.21$\pm$0.06 & $(1.3\pm0.4)\times10^{8}$ & \citet{2024ApJ...962..119L}\\
J1148+5251 & 6.419 & CO(3--2) & 0.20$\pm$0.02 & $(3.8\pm0.4)\times10^{7}$ & \citet{2003Natur.424..406W}\\
 &  & CO(6--5) & 0.67$\pm$0.08 & $(2.5\pm0.3)\times10^{8}$ & \citet{2003AA...409L..47B}\\
 &  & CO(7--6) & 0.63$\pm$0.06 & $(2.8\pm0.3)\times10^{8}$ & \citet{2009ApJ...703.1338R}\\
 &  & CO(17--16) & 0.40$\pm$0.06 & $(4.3\pm0.6)\times10^{8}$ & \citet{2014MNRAS.445.2848G}\\
J0439+1634$^{\dagger}$ & 6.519 & CO(6--5) & 1.5$\pm$0.1 & $(5.8\pm0.4)\times10^{8}$ & \citet{2019ApJ...880..153Y}\\
 &  & CO(7--6) & 1.5$\pm$0.1 & $(6.8\pm0.5)\times10^{8}$ & \citet{2019ApJ...880..153Y}\\
 &  & CO(9--8) & 2.1$\pm$0.2 & $(1.2\pm0.1)\times10^{9}$ & \citet{2019ApJ...880..153Y}\\
 &  & CO(10--9) & 1.9$\pm$0.2 & $(1.2\pm0.1)\times10^{9}$ & \citet{2019ApJ...880..153Y} \\
PJ231$-$20 & 6.586 & CO(7--6) & 0.46$\pm$0.04 & $(2.1\pm0.2)\times10^{8}$ & \citet{2021AA...652A..66P}\\
 &  & CO(10--9) & 0.40$\pm$0.06 & $(2.6\pm0.4)\times10^{8}$ & \citet{2021AA...652A..66P}\\
 &  & CO(15--14) & 0.33$\pm$0.10 & $(3.2\pm1.0)\times10^{8}$ & \citet{2021AA...652A..66P}\\
 &  & CO(16--15) & 0.15$\pm$0.05 & $(1.6\pm0.5)\times10^{8}$ & \citet{2021AA...652A..66P}\\
\enddata
\tablecomments{
$^{\dagger}$ Gravitationally lensed source. Fluxes and luminosities are apparent values.
}
\end{deluxetable*}

\section{Class-Median Normalized CO SLEDs}\label{app:sled_table}

Table~\ref{tab:cosled_norm_band} presents the numerical values of the class-median normalized CO SLEDs plotted in the right panel of Figure~\ref{fig:COSLED_individual+classband}.
The median and 16th--84th percentile range at each integer $J_{\rm up}$ are computed from the PCHIP-interpolated, CO(9--8)-normalized SLEDs described in Section~\ref{sec:results}.

\begin{table}
\centering
\caption{Class-median normalized CO SLEDs relative to CO(9--8).
}
\label{tab:cosled_norm_band}
\begin{tabular}{lccc}
\hline
Transitions & HotDOG & SMG & Quasar \\
\hline
CO(2--1) & $0.02^{+0.01}_{-0.00}$ & $0.04^{+0.16}_{-0.01}$ & $0.04^{+0.04}_{-0.01}$ \\
CO(3--2) & $0.09^{+0.03}_{-0.00}$ & $0.20^{+0.13}_{-0.09}$ & $0.11^{+0.04}_{-0.03}$ \\
CO(4--3) & $0.23^{+0.10}_{-0.03}$ & $0.45^{+0.12}_{-0.18}$ & $0.29^{+0.06}_{-0.11}$ \\
CO(5--4) & $0.40^{+0.13}_{-0.04}$ & $0.75^{+0.17}_{-0.24}$ & $0.47^{+0.22}_{-0.17}$ \\
CO(6--5) & $0.57^{+0.18}_{-0.01}$ & $0.88^{+0.12}_{-0.19}$ & $0.73^{+0.34}_{-0.24}$ \\
CO(7--6) & $0.75^{+0.21}_{-0.03}$ & $0.95^{+0.20}_{-0.14}$ & $0.81^{+0.45}_{-0.22}$ \\
CO(8--7) & $0.91^{+0.08}_{-0.04}$ & $0.98^{+0.13}_{-0.10}$ & $0.93^{+0.25}_{-0.18}$ \\
CO(9--8) & $1.00$ & $1.00$ & $1.00$ \\
CO(10--9) & $1.11^{+0.04}_{-0.07}$ & $1.00^{+0.09}_{-0.11}$ & $1.05^{+0.13}_{-0.17}$ \\
CO(11--10) & $1.18^{+0.28}_{-0.23}$ & $1.01^{+0.29}_{-0.08}$ & $1.12^{+0.19}_{-0.19}$ \\
CO(12--11) & $\ldots$ & $0.97^{+0.12}_{-0.25}$ & $1.18^{+0.25}_{-0.39}$ \\
CO(13--12) & $\ldots$ & $\ldots$ & $1.23^{+0.36}_{-0.31}$ \\
CO(14--13) & $\ldots$ & $\ldots$ & $1.28^{+0.35}_{-0.64}$ \\
CO(15--14) & $\ldots$ & $\ldots$ & $1.31^{+0.70}_{-0.00}$ \\
CO(16--15) & $\ldots$ & $\ldots$ & $1.33^{+0.68}_{-0.51}$ \\
\hline
\end{tabular}
\end{table}

\section{Excitation Parameter and Infrared Luminosities}\label{app:lir_alpha}

Table~\ref{tab:ir_lum_mu} summarizes the excitation parameter $\alpha^\prime$ and the infrared luminosities for the high-redshift sources discussed in Section~\ref{sec:disc_lir_alpha}.
For gravitationally lensed sources, the magnification factor $\mu$ is also listed. 
The adopted $L_{\rm IR}$ values and magnifications are taken from the references given in the table.

\begin{deluxetable*}{lcccl}
\tablewidth{0pt}
\tablecaption{Excitation parameter and infrared luminosity summary for the high-redshift galaxy/quasar sample\label{tab:ir_lum_mu}}
\tablehead{
\colhead{ID} &
\colhead{$\alpha^\prime$} &
\colhead{$\mu$} &
\colhead{$\mu L_{\rm IR}$} &
\colhead{Reference}\\
& & &
($10^{13}~L_\odot$) &
}
\startdata
\multicolumn{5}{c}{Hot DOG sample} \\
\hline
W0116--0505 & $1.49 \pm 0.34$ & --- & 8.2 & \citet{2015ApJ...805...90T} \\
W0831+0140 & $1.73 \pm 0.13$ & --- & 12.0 & \citet{2015ApJ...805...90T} \\
W2246--0526 & $1.52 \pm 0.30$ & --- & 22.1 & \citet{2015ApJ...805...90T} \\
W2246--7143 & $1.54 \pm 0.20$ & --- & 8.3 & \citet{2015ApJ...805...90T} \\
W2305--0039 & $0.90 \pm 0.16$ & --- & 8.3 & \citet{2015ApJ...805...90T} \\
\hline
\multicolumn{5}{c}{SMG sample} \\
\hline
G12v2.30 & $0.68 \pm 0.10$ & 9.5 & $15.7 \pm 1.8$ & \citet{2012ApJ...752..152H,2013ApJ...779...25B} \\
NCv1.143 & $0.94 \pm 0.10$ & 11.3 & $12.8 \pm 4.3$ & \citet{2013ApJ...779...25B,2016AA...595A..80Y} \\
G09v1.97 & $1.29 \pm 0.16$ & 6.9 & $15.3 \pm 4.3$ & \citet{2013ApJ...779...25B,2016AA...595A..80Y} \\
G15v2.779 & $1.12 \pm 0.20$ & 4.6 & $8.6 \pm 0.7$ & \citet{2012ApJ...756..134B,2013ApJ...779...25B} \\
SPT0544$-$40 & $1.12 \pm 0.13$ & 5.5 & $8.1 \pm 5.5$ & \citet{2016ApJ...826..112S,2020ApJ...902...78R} \\
SPT2311$-$54 & $0.93 \pm 0.07$ & 2.0 & $7.7 \pm 1.2$ & \citet{2016ApJ...826..112S,2020ApJ...902...78R} \\
SPT2349$-$56N1 & $1.08 \pm 0.06$ & --- & $1.7 \pm 0.7$ & \citet{2020MNRAS.495.3124H} \\
COS-AzTEC-1 & $1.08 \pm 0.05$ & --- & 1.9 & \citet{2018Natur.560..613T} \\
BR1202$-$0725 & $1.36 \pm 0.14$ & --- & 2.6 & \citet{2006ApJ...645L..97I} \\
COS-AzTEC-3 & $1.08 \pm 0.12$ & --- & $1.1 \pm 0.2$ & \citet{2014ApJ...796...84R} \\
G09-83808 & $1.12 \pm 0.26$ & 8.4 & $3.9 \pm 0.6$ & \citet{2022PASJ...74L...9T} \\
SPT0311$-$58W & $0.98 \pm 0.05$ & 2.1 & $5.4 \pm 2.5$ & \citet{2021ApJ...921...97J} \\
SPT0311$-$58E & $1.17 \pm 0.33$ & 1.3 & $0.46 \pm 0.09$ & \citet{2021ApJ...921...97J} \\
SDP-81 & $0.63 \pm 0.20$ & 11.1 & $5.8\pm0.6$ & \citet{2012ApJ...752..152H,2013ApJ...779...25B} \\
ADFS-27 & $1.08 \pm 0.10$ & --- & $1.6 \pm 0.2$ & \citet{2021ApJ...907...62R} \\
SPT0346--52 & --- & 5.6 & $23.9\pm0.3$ & \citet{2016ApJ...826..112S,2020ApJ...902...78R} \\
HFLS3 & $1.99 \pm 0.60$ & 1.8 & $2.9 \pm 0.3$ & \citet{2020ApJ...895...81R} \\
\hline
\multicolumn{5}{c}{Quasar sample} \\
\hline
BR1202$-$0725 & $0.84 \pm 0.07$ & --- & 1.2 & \citet{2006ApJ...645L..97I} \\
J2310+1855 & $1.34 \pm 0.12$ & --- & $1.2\pm0.4$ & \citet{2022AA...668A.121S} \\
J2219+0102 & --- & --- & 0.18 & \citet{2022AA...662A..60D} \\
J0100+2802 & $1.42 \pm 0.34$ & --- & $0.29\pm0.02$ & \citet{2020ApJ...904..130V} \\
PJ036+03 & $0.98 \pm 0.14$ & --- & $0.58\pm0.01$ & \citet{2020ApJ...904..130V} \\
J2348$-$3054 & $1.04 \pm 0.23$ & --- & $0.57\pm0.02$ & \citet{2020ApJ...904..130V} \\
APM~08279+5255 & --- & 100 & 20 & \citet{2007AA...467..955W} \\
J2322+1944 & --- & 2.5 & 4.0 & \citet{2003Sci...300..773C,2018MNRAS.476.5075S} \\
BRI0952$-$0952 & $1.71 \pm 0.17$ & 3.9 & $3.0\pm1.0$ & \citet{2023AA...673A.116K} \\
J0129$-$0035 & --- & --- & $ 0.48\pm0.01$ & \citet{2020ApJ...904..130V} \\
P215$-$16 & $2.56 \pm 0.37$ & --- & $1.8 \pm 0.2$ & \citet{2024ApJ...962..119L} \\
J2054$-$0005 & $1.42 \pm 0.35$ & --- & $0.62\pm0.02$ & \citet{2020ApJ...904..130V} \\
J1429+5447 & $0.75 \pm 0.18$ & --- & $0.69\pm0.02$ & \citet{2024ApJ...962..119L} \\
J1148+5251 & $1.22 \pm 0.18$ & --- & $2.2 \pm 0.3$ & \citet{2011MNRAS.416.1916V} \\
J0439+1634 & $1.91 \pm 0.17$ & 4.6 & $3.4 \pm 0.2$ & \citet{2019ApJ...880..153Y} \\
PJ231$-$20 & --- & --- & $1.0 \pm 0.03$ & \citet{2020ApJ...904..130V} \\
\hline
\enddata
\end{deluxetable*}

\clearpage

\section{Construction of a GMC-scale shock CO SLED}\label{app:gmc_cshock}

This appendix describes how we synthesize a GMC-scale CO SLED for shocks by combining (i) a turbulent clump population inside a single GMC \citep{2024MNRAS.527.8727E} and (ii) the C-type shock CO line intensities of \citet{2015AA...578A..63F}.
The aim is to compare CO SLED shapes rather than to predict absolute CO luminosities.

Following \citet{2024MNRAS.527.8727E}, we construct three representative GMC models---Tiny, Median, and Huge---with masses of $\log(M_{\rm GMC}/M_\odot)= 3.2$, 4.5, and 5.9, respectively.
For all three models, we assume a fixed surface density $\Sigma_{\rm GMC}=170\,M_\odot\,{\rm pc^{-2}}$.
The GMC radius is therefore
\begin{equation}
R_{\rm GMC}=\left(\frac{M_{\rm GMC}}
{\pi\Sigma_{\rm GMC}}\right)^{1/2},
\end{equation}
and the mean hydrogen number density is
\begin{equation}
n_0=\frac{3M_{\rm GMC}}
{4\pi R_{\rm GMC}^3\,\mu m_{\rm p}},
\end{equation}
where we adopt $\mu=1.22$.

We assume that the logarithmic density contrast $s\equiv \ln(n/n_0)$ follows a lognormal distribution whose variance is set by supersonic turbulence:
\begin{equation}
\sigma_s^2=\ln\!\left(1+b^2\mathcal{M}^2\right),
\end{equation}
where $\mathcal{M}$ is the sonic Mach number and $b$ is the turbulence forcing parameter.
We adopt $\mathcal{M}=10$ and $b=0.5$.
Discrete clumps are generated by drawing $s$ from a Gaussian with dispersion $\sigma_s$, retaining only draws with $s\ge 0$.
We employ a mass-weighted lognormal, corresponding to a mean of $s_0=+\sigma_s^2/2$, as opposed to the volume-weighted case $s_0=-\sigma_s^2/2$.
This choice preferentially samples the dense structures that dominate the GMC mass and potentially the shock-powered CO emission.

\begin{figure}[t]
  \centering
  \includegraphics[scale=1.0]
  {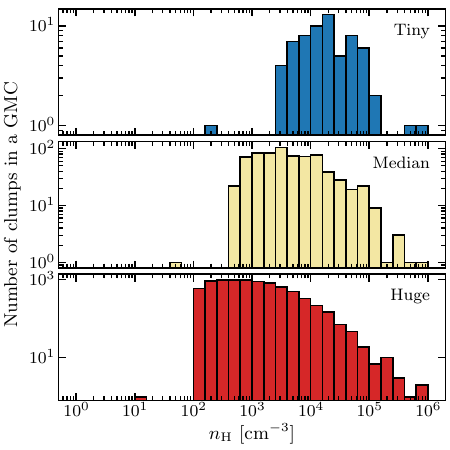}
  \caption{
Histogram of the hydrogen number density of synthetic clumps extracted within three GMC models.
Tiny, Median, and Huge correspond to GMC masses of $\log(M_{\rm GMC}/M_\odot)= 3.2$, 4.5, and 5.9, respectively \citep{2024MNRAS.527.8727E}.
}
  \label{fig:app_clump_hist}
\end{figure}

Each accepted density draw is assigned a characteristic clump radius based on a Jeans-length-like scale,
\begin{equation}
R_{\rm cl}(n)=f_{\rm RJ}\left(\frac{\pi c_s^2}
{G\rho}\right)^{1/2},
\end{equation}
where $\rho=\mu m_{\rm p}n$ and the sound speed is $c_s=\left(\gamma k_{\rm B}T/\mu m_{\rm p}\right)^{1/2}$.
We adopt $T=10$~K and $\gamma=5/3$.
The factor $f_{\rm RJ}$ is a scaling parameter that accounts for uncertainties in the characteristic clump size.
We adopt $f_{\rm RJ}=0.5$, which yields clump density histograms broadly consistent with the published clump distributions in \citet{2024MNRAS.527.8727E} for comparable GMCs (Figure~\ref{fig:app_clump_hist}).

\citet{2015AA...578A..63F} provide integrated intensities $W_J$ (in units of K\,km\,s$^{-1}$) for CO rotational transitions in C-type shocks at discrete pre-shock densities and shock velocities.
We consider three fixed shock velocities, $v_{\rm s}=10$, 20, and 30~km~s$^{-1}$, and use the density grid available in the catalog ($\log n/{\rm cm^{-3}} = 3,4,5,6$).
For each clump density $n_i$, we obtain $W_J(n_i,v_{\rm s})$ by linear interpolation in $\log n$ between the two nearest grid points, with clipping applied at the grid boundaries to avoid extrapolation.

To construct a GMC-scale CO SLED, we convert $W_J$ into a luminosity-like proxy for each clump.
Assuming that the shock emissivity is specified per unit surface area and that the effective emitting area of each clump scales as its surface area, $A_i\propto 4\pi R_{{\rm cl},i}^2$, the line luminosity proxy can be written as
\begin{equation}
L_{J}^{\rm (GMC)}(v_{\rm s}) \propto
\sum_i A_i\,\nu_J^3\,W_J(n_i,v_{\rm s}),
\end{equation}
where the $\nu_J^3$ factor accounts for the Rayleigh--Jeans conversion between brightness temperature and line energy output.
We then normalize the resulting CO SLED to the CO(9--8) transition.

\begin{figure}
  \centering
  \includegraphics[scale=1.0]
  {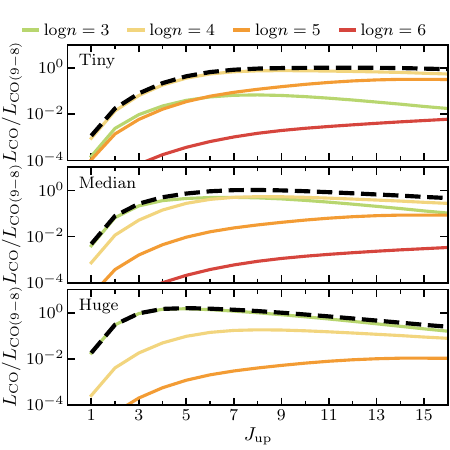}
  \caption{
Density-decomposed GMC-scale CO SLEDs at a fixed shock velocity of $v_{\rm s}=30~\mathrm{km~s^{-1}}$.
From top to bottom, panels show the three GMC models with $M_{\rm GMC}=10^{3.2}$, $10^{4.5}$, and $10^{5.9}\,M_\odot$, respectively.
The black dashed curve shows the total CO SLED obtained by summing the shock emission over all clumps.
The colored curves show the partial SLEDs from clumps grouped by pre-shock gas density.
}
  \label{fig:COSLED_clump_contrib}
\end{figure}

Figure~\ref{fig:COSLED_clump_contrib} visualizes how the GMC-scale CO SLED is built up from clumps spanning the internal density distribution.
Because we adopt a mass-weighted sampling, most of the GMC mass resides at moderate densities, while the high-density tail contains only a small fraction of the total mass.
As a result, the GMC-integrated CO SLED is dominated by contributions from clumps with $n\sim10^{3}$--$10^{4}~\mathrm{cm^{-3}}$.
At fixed surface density, increasing the GMC mass increases the cloud radius and decreases the characteristic mean density, shifting the dominant contribution toward lower-density clumps and producing a less excited CO SLED.

\begin{figure}
  \centering
  \includegraphics[scale=1.0]
  {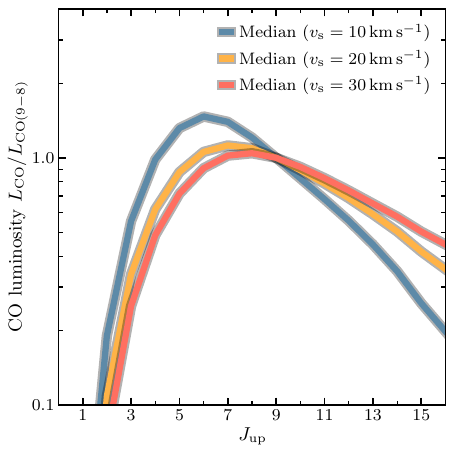}
  \caption{
GMC-scale CO SLEDs for shocks.
Curves show fixed shock velocities $v_{\rm s}=10$, 20, and 30~km~s$^{-1}$ for a GMC model with $\log(M_{\rm GMC}/M_\odot)=4.5$.
}
  \label{fig:app_gmcshock_sled}
\end{figure}

Figure~\ref{fig:app_gmcshock_sled} presents the resulting GMC-scale CO SLEDs for $v_{\rm s}=10$, 20, and 30~km~s$^{-1}$ for a representative GMC with $\log(M_{\rm GMC}/M_\odot)=4.5$.
These models do not maintain a strongly excited, flat high-$J$ tail even at a relatively fast shock velocity ($v_{\rm s}=30~\mathrm{km~s^{-1}}$).

\clearpage

\bibliographystyle{apj}
\bibliography{ref}



\end{document}